\documentclass{SciPost}
\usepackage{physics}
\usepackage{graphicx} 
\usepackage{subcaption}
\usepackage{float}

\begin{document}

\begin{center}{\Large \textbf{
Density of states correlations in L\'evy Rosenzweig-Porter model via supersymmetry approach
}}\end{center}

\begin{center}
Elizaveta Safonova\textsuperscript{1,2}, 
Aleksey Lunkin\textsuperscript{1} and 
Mikhail Feigel' man\textsuperscript{1,3}
\end{center}

\begin{center}
{\bf 1} Nanocenter CENN, Ljubljana, Slovenia
\\
{\bf 2} Faculty of Mathematics and Physics, University of Ljubljana, Jadranska 19, 1000 Ljubljana, Slovenia\\
{\bf 3} Jozef Stefan Institute, Ljubljana, Slovenia
\end{center}

\begin{center}
\today
\end{center}


\section*{Abstract}
{\bf 
We studied global density-of-states correlation function 
$R(\omega)$ for L\'evy-Rosenzweig-Porter random matrix ensemble~\cite{BirTar_Levy-RP} in the non-ergodic extended phase.  Using an extension of Efetov's supersymmetry  approach ~\cite{MirFyod_non_Gauss} we calculated 
$R(\omega)$ exactly  in all relevant  ranges of $\omega$. 
At relatively 
low $\omega \leq \Gamma$\, (with $\Gamma \gg \Delta$ being the effective miniband width) we found GUE-type oscillations with period of level spacing $\Delta$, decaying exponentially 
at the Thouless energy scale $E_{Th} = \sqrt{\Delta \Gamma/2\pi}$. At high energies $\omega \gg E_{Th}$ our results coincide with
those obtained in Ref.~\cite{lunkin2024localdensitystatescorrelations} via cavity equation approach. Inverse of the effective miniband width, $1/\Gamma$, is shown to be given by the average of the local decay times over L\'evy distribution.
}

\vspace{10pt}
\noindent\rule{\textwidth}{1pt}
\tableofcontents\thispagestyle{fancy}
\noindent\rule{\textwidth}{1pt}
\vspace{10pt}
\section{Introduction}
\label{sec:intro} 
There are numerous indications for the apparent absence of thermalization and breakdown of ergodicity  in large interacting  quantum systems~\cite{RevModPhys.91.021001,Vidmar-review,PhysRevLett.120.050507} with sufficiently high degree of disorder. However, almost no exact theoretical results are available, making reliable interpretation of real and numerical experiments rather complicated. While the original theoretical approach to this problem~\cite{BAA,Mirlin_MBL} was focused on low-temperature transport properties, later development in this field (now called the Many Body Localization (MBL) problem)  was shifted mainly to the infinite-temperature limit, for the sake of simplification; also, some types of experiments (NMR, cold atoms) may indeed be realized at effective temperatures much higher than typical energies involved in the Hamiltonian.  Still the issue of existence of non-ergodic and/or MBL state in a real physical system with short-range interaction is highly debatable~\cite{PhysRevE.102.062144,PhysRevB.105.224203}.

One of the major obstacles for the theory of MBL phenomena is the presence of well-developed spatial correlations.  Indeed, while dimension of Hilbert space of a random system containing $n$ spins-$\frac12$ is $2^n$, the number of parameters entering its Hamiltonian is just $\sim n^2$ at most.
Proper account of these correlations is not developed yet, and theoretical results are limited to some artificial models where these correlations are absent.  In particular, it was shown in Ref.~\cite{QREM2019} that structureless Quantum Random Energy Model possesses three different phases,
depending on macroscopic energy and degree of disorder: ergodic, fully localized (MBL) and intermediate non-ergodic extended (NEE)  state.
Theoretical demonstrations of these features were obtained by means of approximate mapping of the QREM Hamiltonian to the Rosenzweig-Porter matrix model shown previously~\cite{gRP} to have all three such phases.  It was understood later on~\cite{KhayKrav_LN-RP,BirTar_Levy-RP} that Gaussian RP model~\cite{gRP} is oversimplified to describe more realistic problems; one possible way to generalize this model is to account for the possibility of fat-tailed distribution of non-diagonal matrix elements. An independent reason to be interested in this kind of models is due to (numerical) observations of a power-law distribution of matrix elements connecting different bit-strings in systems of interacting  quantum 
spins~\cite{Long2023,Roy-Logan_2020_corr,QIsing_2020}.

Invariant L\'evy matrix ensemble was introduced originally in Ref.~\cite{Bouchard_Levy_Mat} and its Rosenzweig-Porter version was studied in Ref.~\cite{BirTar_Levy-RP,lunkin2024localdensitystatescorrelations}. In particular, Ref.~\cite{BirTar_Levy-RP} demonstrated the presence of NEE state in the whole range of parameters $\mu,\gamma$ characterizing the model, while in  Ref.\cite{lunkin2024localdensitystatescorrelations} full description of local density-of-states correlations at large energy difference (effectively, setting level spacing to zero) was obtained by means of statistical analysis of cavity equations.  However, to study level correlations at low energy difference
$\omega \leq \Delta \sim 1/N$, a more elaborate technique is needed. Indeed,  cavity equation approach is valid in the $N \to \infty$ limit, equivalent to $\Delta \to 0$. 

Well-developed methods to treat this type of problem in usual random-matrix ensembles are based on the supersymmetry method due to Efetov~\cite{Efetov_book}. Application of this method to Gaussian RP model was recently provided in Ref.~\cite{Skv-Krv}.  However, standard SUSY method based upon Hubbard-Stratonovich transformation of the functional integral is not appropriate for matrix models with a heavy-tailed distributions, especially when second moment of the distribution diverges, as in the L\'evy case.  More general approach to the construction of supersymmetric field theory for disordered quantum systems was proposed in Ref.~\cite{MirFyod_non_Gauss}, where functional generalization of the Hubbard-Stratonovich transformation was introduced.  In the previous paper~\cite{SafonovaLevyRPDoS} we employed this approach to study the average density of states of L\'evy-Rosenzweig-Porter  ensemble.  Below we extend this approach for the calculation of the global density-of-states correlation function $R(\omega) = \langle \rho(E+\omega/2)\rho(E-\omega/2)\rangle / \langle \rho(E)\rangle^2$ at arbitrary $\omega$ in the NEE state.  We demonstrate the presence of three energy scales in the problem: mean level spacing $\Delta$,  typical miniband width $\Gamma \gg \Delta$ and intermediate scale 
$E_{Th} = \sqrt{\Gamma\Delta/2\pi}$  which plays the role of a Thouless energy in our problem, similar to the result of Ref.~\cite{Skv-Krv} for Gaussian RP model, see also~\cite{Tomasi-Krav}.  
Previous results~\cite{lunkin2024localdensitystatescorrelations} are confirmed for $\omega \gg E_{Th}$ by our supersymmetry method, while at low  $\omega \leq E_{Th}$ the function $R(\omega)$ demonstrates oscillations typical for Wigner-Dyson random matrix ensembles.


\begin{figure}[h!]
    \centering
    \includegraphics[width=0.5\linewidth]{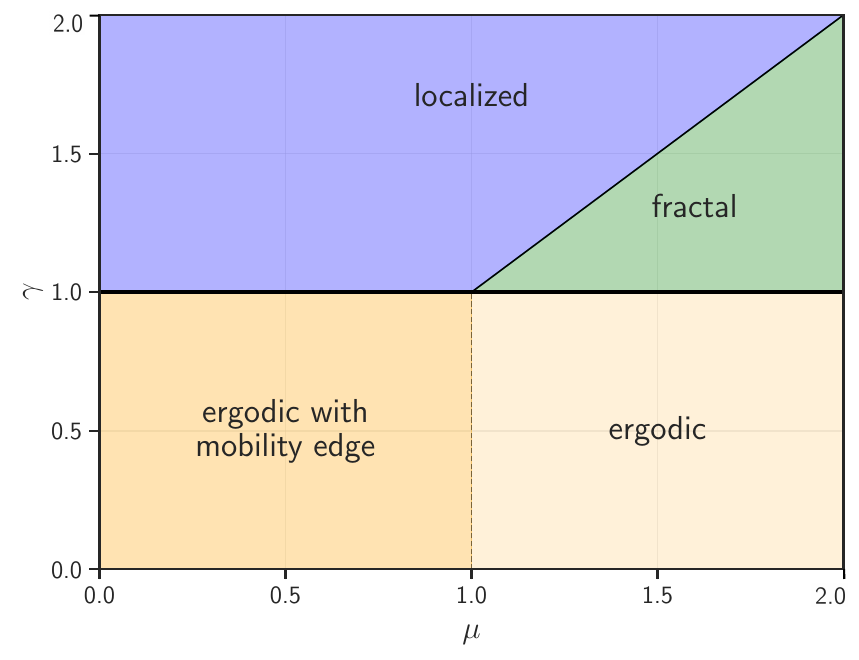}
    \caption{Different regimes depending on the width
of miniband $\Gamma_{0}$ in comparison with level spacing; boundaries between them are in agreement with results
of Ref.~\cite{BirTar_Levy-RP}. $\Gamma_{0}$ depends on the system
size $N$ and the parameters $\gamma,\mu$ as 
$\sim N^{\frac{1-\gamma}{\mu-1}}$ and determines behavior of the system (See \eqref{eq:=000020Gamma0=000020def} and comment under it). 
For $\gamma < 1$ the system is ergodic; if $\mu < 1$, then there is
also mobility edge (transition to localized states at energies
closer to the band edge). We are interested in the range of
$\gamma > 1$ where eigenstates are either localized ($\gamma > \mu -1$) or extended but non-ergodic at $\gamma < \mu -1$. In this latter phase $\Gamma_0 $ is much larger than level spacing but much smaller than with whole bandwidth $W$.}
    \label{fig:phase-diag}
\end{figure}

Before going into the calculations, we briefly review the main features of the phase diagram for L\'evy-RP matrices, based mostly on Ref.\cite{BirTar_Levy-RP}. The part of the phase diagram we’re interested in covers the range $1 < \mu < 2$, and it’s shown in Fig.\eqref{fig:phase-diag}. The different phases are defined based on the behavior of the eigenvectors $\Psi_n(i)$. These can be ergodic, where the inverse participation ratio (IPR, $I(N) = \sum_n^N |\Psi_n|^4$) scales like $I(N) \sim N^{-1}$, localized with $I(N) \sim$ constant, or non-ergodic but extended — with $I(N) \sim N^{-D}$ for some $0 < D < 1$.
There are two ergodic (E) phases. One appears for $1 < \mu < 2$, where all eigenvectors are ergodic for any energy $E_n$. The other is for $0 < \mu < 1$, where a mobility edge $E_0$ separates ergodic and localized states: eigenvectors are ergodic when $|E_n| < E_0$ and localized when $|E_n| > E_0$. All three phases—ergodic, localized, and non-ergodic extended—meet at the tricritical point $\mu = \gamma = 1$.

In this paper we are concerned with the correlation function defined by Eq.\eqref{correlator=000020def} at $1<\mu<2$. We show that Lévy-RP model indeed experiences phase transitions at the boundaries $\gamma = \mu$ and $\gamma = 1$ and we provide the explicit analytical calculation.

The rest of the paper is organized as follows. Sec.2 introduces definitions of the random matrix ensemble we are going to study and
representation of the correlation function $R(\omega)$ in terms of functional integral over superfields.  Sec. 3 describes functional Hubbard-Stratonovich transformation and provides saddle-point analysis  of the relevant functional integral. Sec. 4 is devoted to calculation of the correlation function $R(\omega)$  in two overlapping limiting cases:  high frequencies $\omega \gg E_{Th}$ and low frequencies $\omega \ll \Gamma$.  Since $E_{Th} \ll \Gamma$,  we thus obtain the full behavior of $R(\omega)$ in the whole range of frequencies. Sec. 5 contains our conclusions. Supplemental material (Secs. A-F) contains technical details of our calculations.

\section{Definitions}

\subsection{The matrix ensemble}

Our research object is an ensemble of
$N\times N$ complex Hermitian matrix $\hat{H}$
which can be represented as the sum of two matrices: 
\begin{equation}
\hat{H}=\hat{H}_{D}+\hat{H}_{L},
\label{H0}
\end{equation}
where $\hat{H}_{D}$ is a diagonal random matrix with real independent
and identically distributed (i.i.d.) entries and $\hat{H}_{L}$ is
a full matrix where \textit{all} elements are i.i.d. The distributions
of $\hat{H}_{L}$ and $\hat{H}_{D}$ are generally different. We consider
the case of the \textit{Lévy-Rosenzweig-Porter (Lévy-RP) matrices}
\cite{BirTar_Levy-RP} where the entries of $\hat{H}_{D}$ are random,
broadly distributed with the typical distribution width $W$ so that
$W$ is the largest energy scale. 
Level spacing $\Delta \sim W/N$ is the smallest energy scale.
$H_{L}$ entries are complex and defined as follows: 
\begin{equation}
[H_{L}]_{mn}=h_{mn}\exp\left(i\theta_{mn}\right),\quad h_{nm}\geq0,\quad-\pi\leq\theta_{mn}<\pi.
\label{matrix def}
\end{equation}
The phase $\theta_{nm}$ is distributed uniformly with $P_{\theta}\left(\theta\right)=\frac{\theta\left(\pi-\left|\theta\right|\right)}{2\pi}$
and the amplitudes $h_{mn}$ have a distribution according to the
power-law. For convenience, we chose the particular one-sided Lévy
distribution

\begin{equation}
P_{L}^{(\mu,\gamma)}\left(h_{mn}^{2}\right)=\frac{N^{\frac{2\gamma}{\mu}}}{\sigma^{2}}L_{\mu/2}\left(\frac{N^{\frac{2\gamma}{\mu}}}{\sigma^{2}}h_{mn}^{2}\right),\label{offdiag=000020dist}
\end{equation}
where $\sigma$ is an energy unit and $L_{\mu/2}(x)$ is \textit{one-sided}
Lévy stable distribution \cite{Levy_fr,Mandelbrot} which is defined
by Laplace characteristic function: 
\begin{equation}
\tilde{L}_{\mu/2}(r)\equiv\int_{0}^{\infty}L_{\mu/2}(x)e^{-rx}dx\equiv e^{-r^{\mu/2}},\quad1<\mu\leq2.\label{char-fun}
\end{equation}
We chose that particular function because it supports only positive
values and has a convenient representation in terms of its Laplace
transform. Using Eqs.(\ref{char-fun}),(\ref{offdiag=000020dist})
 the Laplace characteristic function of rescaled $P_{L}$
distribution:

\begin{equation}
\int_{0}^{\infty}P_{L}\left(h^{2}\right)e^{-rh^{2}}d\left[h^{2}\right]\equiv\exp\left(-\frac{\sigma^\mu}{N^{\gamma}}r^{\mu/2}\right),\quad\begin{array}{c}
1<\mu<2\\
\gamma>0
\end{array}
\end{equation}
In fact, any distribution with the same power-law tail 
will lead to  similar results, as explained in the end of the paper. Function (\ref{offdiag=000020dist}) has the following
power-law asymptotics
\begin{equation}
P_{L}\left(h^{2}\right)dh^{2}\approx\frac{\mu\sigma^{\mu}dh}{\Gamma\left(1-\frac{\mu}{2}\right)N^{\gamma}h^{1+\mu}},\quad1<\mu<2\label{Levy=000020asymptotic}
\end{equation}
For $\mu\geq2$ this distribution has a finite variance and the model
becomes equivalent to the usual Gaussian Rosenzweig-Porter model.
To compare intermediate results with the previous papers
\cite{Skv-Krv}, \cite{Biroli_Levy_Mat}, one can put $\sigma=1$,
while notations of Ref.~\cite{lunkin2024localdensitystatescorrelations}
are recovered by the choice $\gamma=1$ and  $\quad\frac{\sigma^{\mu}}{\Gamma\left(1-\frac{\mu}{2}\right)}=h_{0}^{\mu}$.

Note that while the variance $W^{2}$ of $H_{D}$ is independent
of the matrix size $N$, the typical value of $H_{L}$ scales with
$N$ as $\sigma N^{-\gamma/\mu}$, and its variance diverges at $\mu<2$
due to the tail in $L_{\frac{\mu}{2}}(x^{2})\sim x^{-(1+\mu)}$. There
is a special value $\mu=2$ where the distribution $L_{\mu/2}(x)$
reduces to the delta function $\delta(x-1)$.

\subsection{Global DoS correlation function and supersymmetric method}

Our goal  is to calculate correlation function of global density of states which is defined as
\begin{equation}
R\left(E,\omega\right)=\frac{\left\langle \rho\left(E+\frac{\omega}{2}\right)\rho\left(E-\frac{\omega}{2}\right)\right\rangle }{\left\langle \rho\left(E\right)\right\rangle ^{2}}\label{correlator=000020def}
\end{equation}
where $\rho(E) = -\frac1{\pi N}\text{Tr}\Im \hat{G}_R(E)$ is density of states (DoS) and $\hat{G}_R (E)$ is retarded Green function of the Hamiltonian (\ref{H0}) at energy $E$. It is convenient to choose
the scaling so that DoS becomes a function of the order of unity:
\begin{equation}
\frac{1}{\Delta}=N\rho\left(E\right),\qquad\int dE\rho\left(E\right) = 1
\end{equation}
where $\Delta$ is mean level spacing. To continue the calculation
one should switch to the Green function representation, so that the
correlation function is 
\begin{equation}
R\left(E,\omega\right)=\frac{1}{2}+\frac{\Delta^{2}}{2\pi^{2}}\text{Re}\left\langle \text{Tr}\hat{G}_{R}\left(E+\frac{\omega}{2}\right)\text{Tr}\hat{G}_{A}\left(E-\frac{\omega}{2}\right)\right\rangle \label{correlation=000020function=000020Green=000020representation}
\end{equation}

Two-point correlation function can be expressed through differentiation
the partition function $Z(E,\omega,J_{A},J_{R})$ over background
fields $J_{R},J_{A}$ . The partition function $Z(E,\omega,J_{A},J_{R})$
is given~\cite{Efetov_book,Mirlin2000} by the integral over supervectors $\psi_{i}$ (for the derivation, see Supplement, Sec.\ref{A2}).
\begin{equation}
R\left(E,\omega\right)=\frac{1}{2}+\frac{\Delta^{2}}{8\pi^{2}}\text{Re}\frac{\partial^{2}Z\left(E,\omega,\hat{J}\right)}{\partial J_{R}\partial J_{A}}\biggr|_{J_{R,A}=0}\label{correlator=000020with=000020sources}
\end{equation}
\begin{equation}
Z\left(E,\omega,\hat{J}\right)=\left\langle \int\left[d\psi\right]\exp\left(i\sum_{n,m}\psi_{n}^{\dagger}\hat{L}\left(\left[E+\frac{\Omega}{2}\hat{L}-\hat{J}\hat{K}\right]\delta_{nm}-H_{nm}\right)\psi_{m}\right)\right\rangle _{\hat{H}}\label{partitio=000020function}
\end{equation}
where $\Omega\equiv\omega+i0$ (here and below infinitesimal imaginary part is introduces to guarantee convergence of the integrals). Expression (\ref{partitio=000020function}) uses superalgebra
formalism which includes commuting and anticommuting variables:
\begin{equation}
\psi_{i}=\left(\begin{array}{c}
\psi_{R}\\
\psi_{A}
\end{array}\right)=\left(\begin{array}{c}
S_{i1}\\
\chi_{i1}\\
S_{i2}\\
\chi_{i2}
\end{array}\right),\quad\psi_{i}^{\dagger}=\left(\begin{array}{cc}
\psi_{R}^{\dagger} & \psi_{A}^{\dagger}\end{array}\right)=\left(\begin{array}{cccc}
S_{i1}^{*} & \chi_{i1}^{*} & S_{i2}^{*} & \chi_{i2}^{*}\end{array}\right)
\end{equation}
are 4-dimensional supervectors with ordinary (complex, commuting)
$\left(S_{i1},S_{i2}\right)$ and Grassmanian (anticommuting) $\left(\chi_{i1},\chi_{i2}\right)$
components, 
\begin{equation}
\hat{K}=\left(\begin{array}{cc}
1\\
 & -1
\end{array}\right)_{BF}=\text{diag}\left(\begin{array}{cccc}
1 & -1 & 1 & -1\end{array}\right),
\end{equation}
\begin{equation}
\hat{L}=\left(\begin{array}{cc}
1\\
 & -1
\end{array}\right)_{RA}=\text{diag}\left(\begin{array}{cccc}
1 & 1 & -1 & -1\end{array}\right),
\end{equation}
\begin{equation}
\hat{J}=\left(\begin{array}{cc}
J_{R}\\
 & J_{A}
\end{array}\right)_{RA}=\text{diag}\left(\begin{array}{cccc}
J_{R} & J_{R} & J_{A} & J_{A}\end{array}\right)
\end{equation}

and $\left[d\psi\right]=\left[d\psi_{R}d\psi_{R}^{\dagger}\right]\left[d\psi_{A}d\psi_{A}^{\dagger}\right]$. 

\section{Functional integral and saddle point equations}

The goal of this section is to derive a  proper field theory ($\sigma$-model)  which describes energy level correlations in the system at sufficiently low energies $ \sim \Delta$.
Starting from Eq.(\ref{partitio=000020function}) one needs to perform
quite a number of mathematical calculations, which are described 
in detail in Section~\ref{B} of the Supplement. To put it briefly, the first step is to average over realizations of 
Lévy distributed matrix elements.
Next step in a usual supersymmetric approach is to use
Hubbard-Stratonovich transformation, which however cannot be used in our
case of the power-law tailed distributions, since its second moment $\overline {|H_{ij}|^2}$
diverges (while it must be the crucial parameter within the standard scheme~\cite{Efetov_book,Mirlin2000,Mirlin:2000cla}).
Instead, we use 
functional analogue of the Hubbard-Stratonovich transformation, which was proposed and described in detail in {[}\cite{MirFyod_non_Gauss}, see
also~\cite{SafonovaLevyRPDoS}{]}. Following this approach (see also Sec.~\ref{B}2), the partition function (\ref{partitio=000020function}) can be transformed into the following functional integral over functions 
$g(\psi,\psi^+)$ dependent on supervectors $\psi$ and $\psi^+$: 
\begin{equation}
Z\left(E,\omega,\hat{J}\right)=\int Dg\exp\left(S\left[g\left(\psi,\psi^{\dagger}\right)\right]\right),\label{eq:=000020Z=000020=00003D=000020int=000020dg=000020e(S=00005Bg=00005D)}
\end{equation}
where the functional action $S\left[g\left(\psi,\psi^{\dagger}\right)\right]$
is given by
\begin{equation}
S\left[g\left(\psi,\psi^{\dagger}\right)\right]=N\ln\left\langle \int\left[d\psi\right]\exp\left(i\psi^{\dagger}\left(E\hat{L}+\frac{\Omega}{2}-\hat{J}\hat{K}\hat{L}-\zeta\hat{L}\right)\psi-g\left(\psi,\psi^{\dagger}\right)\right)\right\rangle _{\zeta}+\label{Action=000020generalized}
\end{equation}
\[
\frac{N}{2}\int\left[d\psi\right]\left[d\psi'\right]g\left(\psi,\psi^{\dagger}\right){\cal I}^{-1}\left(\psi^{\prime\dagger}\hat{L}\psi\right)g\left(\psi^{\prime},\psi^{\prime\dagger}\right).
\]
where ${\cal I}\left(x\right)\equiv\frac{\sigma^{\mu}N^{1-\gamma}}{\Gamma\left(\frac{\mu}{2}+1\right)}\left[x^{\dagger}x\right]^{\mu/2}$
and $\zeta$ corresponds to diagonal elements. 
Variable $\zeta$ stays for elements of diagonal
matrix $H_{D}$ and its distribution is smooth at the scale
of bandwidth $W$.

Due to the large prefactor $N$ in the action, one can perform the
functional integration over $g(\psi,\psi^+)$ by the steepest descent method which leads to the self-consistency equation, whose explicit form
depends on the energy argument $\omega$: 
\begin{equation}
g_{\omega}\left(\psi^{\prime\dagger}\psi^{\prime},\psi^{\prime\dagger}\hat{L}\psi^{\prime}\right)=\left\langle \int\left[d\psi\right]{\cal I}\left(\psi^{\prime\dagger}\hat{L}\psi\right)\exp\left(i\psi^{\dagger}\left(E\hat{L}-\zeta\hat{L}+\frac{\Omega}{2}\right)\psi-g_{\omega}\left(\psi^{\dagger}\psi,\psi^{\dagger}\hat{L}\psi\right)\right)\right\rangle _{\zeta}
\label{s.p.=000020omega}
\end{equation}
As follows from the form of Eq.(\ref{s.p.=000020omega}), its solution 
depends on two invariant objects:
$\psi^{\prime\dagger}\psi^{\prime}$ and 
$\psi^{\prime\dagger}\hat{L}\psi^{\prime}$. Details of the solution
of Eq.(\ref{s.p.=000020omega}) are provided in Sec.\ref{C} of the Supplement.

This self-consistency equation is, in fact, the equation for the cumulant-generating function of the joint distribution of the real and imaginary parts of the self-energy \cite{FyodorovMirlinSommers}. It is equivalent to the equation obtained by the cavity method in the limit $N \to \infty$ \cite{lunkin2024localdensitystatescorrelations}. Nevertheless, for the subsequent analysis it is instructive to derive and solve this equation explicitly within our present method.

The key physical observation which helps to solve Eq.(\ref{s.p.=000020omega}) 
goes as follows:
$e^{-g_{\omega}\left(\psi^{\dagger}\psi,\psi^{\dagger}\hat{L}\psi\right)}$ is the characteristic function of a complex self-energy function $\Sigma$ of the operator $(E - \hat{H})^{-1}$.
Reduced functions of only single arguments,
$e^{-g_{\omega}\left(0,\psi^{\dagger}\hat{L}\psi\right)}$
and $e^{-g_{\omega}\left(\psi^{\dagger}\psi,0\right)}$, 
represent characteristic functions of real and imaginary part of the
same self-energy, respectively. 
Now, the key point is that the full function $g_{\omega}\left(\psi^{\dagger}\psi,\psi^{\dagger}\hat{L}\psi\right) $
can be represented as a simple sum of two independent functions:
\begin{equation}
g_{\omega}\left(\psi^{\dagger}\psi,\psi^{\dagger}\hat{L}\psi\right)
\approx 
g_{\omega}\left(0,\psi^{\dagger}\hat{L}\psi\right)+
g_{\omega}\left(\psi^{\dagger}\psi,0\right)
\label{eq:=000020independency=000020g}
\end{equation}
It means that real and imaginary parts of the self-energy $\Sigma$ are independently distributed. Physical reason for such an independence
is that $\Re\Sigma(E)$ acquires relevant contributions from a 
broad range of energies $E \sim W$, while $\Im\Sigma(E)$ is determined
by the close vicinity of $E$ only. This phenomenon was studied in detail
in Ref.~\cite{lunkin2024localdensitystatescorrelations}.
The distribution of $\Re\Sigma$ was evaluated in our previous paper~\cite{SafonovaLevyRPDoS} where the average density of states was calculated. It leads to a slight renormalization of  spectrum $\sim\frac{\sigma}{W}$ and can be omitted in the present problem. 
The reason can be seen in Eq.(\ref{s.p.=000020omega}):
integration over $d\zeta$ over the broad range $\sim W$ makes very
small relevant values of $\psi^{\dagger}\hat{L}\psi\leq\frac{1}{W}$.
As a result, it is sufficient to work with 
$g_{\omega}\left(\psi^{\dagger}\psi,0\right)$.


At sufficiently large $\omega$ saddle-point solution of the type of
(\ref{eq:=000020independency=000020g}) is sufficient for the purpose of
our calculations (precise criterion on the range of $\omega$ will be
present below). The corresponding solution is described in Sec.~\ref{C}
of the Supplement, the result  reads a follows:
\begin{equation}
g_{\text{s.p.}}\left(\psi,\psi^{\dagger}\right)\biggr|_{\psi^{\dagger}\hat{L}\psi=0}=g_{\omega}\left(\psi^{\dagger}\psi,0\right)=\left[\Gamma_{\omega}\psi^{\dagger}\psi\right]^{\mu/2}
\label{eq:=000020saddle=000020solutions=000020g}
\end{equation}
where function $\Gamma_{\omega}$ is determined by the transcendental equation 
\begin{equation}
\Gamma_{\omega}^{\mu-1}=\Gamma_{0}^{\mu-1}\frac{\Gamma\left(\frac{\mu}{2}\right)}{\Gamma\left(2-\frac{2}{\mu}\right)}\int_{0}^{\infty}drL_{\mu/2}\left(r\right)\left[r-i\frac{\omega}{\Gamma_{\omega}}\right]^{1-\frac{\mu}{2}},
\label{Gomega}
\end{equation}
and its zero-frequency limit $\Gamma_0$ is  expressed via energy parameters
$\sigma$ and $\Delta$ as follows:
\begin{equation}
\left[\frac{\Gamma_{0}}{2}\right]^{\mu-1}=\frac{\sigma^{\mu}}{\Delta N^{\gamma}}\frac{\sqrt{\pi}\Gamma\left(\frac{\mu-1}{2}\right)\Gamma\left(2-\frac{2}{\mu}\right)}{\Gamma^{2}\left(\frac{\mu}{2}\right)}
\label{eq:=000020Gamma0=000020def}
\end{equation}
with $\Gamma(x)$ in the R.H.S. being Euler Gamma-functions.

To meet the requirements of intermediate non-ergodic state one needs to apply the constraint $\Delta \ll \Gamma_0 \ll W$ in $N\rightarrow \infty$ limit (otherwise saddle point approximation is not valid). This will lead to inequalities: $N^{\frac{\gamma}{\mu} - 1} < \frac{\sigma}{W} < N^{\frac{\gamma - 1}{\mu}}$. However, numerical prefactor in \eqref{eq:=000020Gamma0=000020def} strongly diverges at $\mu \rightarrow 1$ so one should be careful with the choice of specific parameters while doing numerical study.


Few remarks are in order now.  First, we note that the form of the saddle-point solution (\ref{eq:=000020saddle=000020solutions=000020g})
demonstrates a heavy-tail nature of distributions of $\Im\Sigma$ and $\Im G$. Second, we emphasize the appearance of a new energy scale
$\Gamma_0$ determined by Eq.(\ref{eq:=000020Gamma0=000020def}),
see also Ref.~\cite{BirTar_Levy-RP}. In the Gaussian case $\mu =2$ it gives just the value of the miniband
width~\cite{Skv-Krv}, while for generic $1< \mu <2$  miniband structure is more complicated, it is characterized by a distribution of
widths which is characterized by the parameter given by  Eq.(\ref{eq:=000020Gamma0=000020def}); the same equation for $\Gamma_0$ was obtained
in Ref.~\cite{lunkin2024localdensitystatescorrelations}.  Third,  at nonzero $\omega$ the function $\Gamma_\omega$ is complex, with
$\Im \Gamma_\omega < 0$; this feature is related to the analytic properties of the DoS correlation function and it will be important later in Sec.~\ref{sec: answers}.




At high  frequencies $\omega \geq \Gamma_0$ the function $\Gamma_\omega$  can be obtained from Eq.(\ref{Gomega}) and behaves as
\begin{equation}
\frac{\Gamma_{\omega\rightarrow\infty}}{\Gamma_{0}}\sim\left|\frac{\omega}{\Gamma_{0}}\right|^{\frac{2}{\mu}-1}
\left(\cos \left[\frac{\pi(2-\mu)}{2\mu}\right] - i
\sin\left[\frac{\pi(2-\mu)}{2\mu}\right]
\right)
\left[\frac{\Gamma\left(\frac{\mu}{2}\right)}{\Gamma\left(2-\frac{2}{\mu}\right)}\right]^{\frac{2}{\mu}}.
\end{equation}
Since L\'evy distribution degenerates into a delta function at $\mu=2$,
$\Gamma_{\omega}$ becomes real constant $\Gamma_{\omega}=\Gamma_{0}$ regardless of $\omega$.
On the other hand, at $\omega=0$ saddle-point solution (\ref{eq:=000020saddle=000020solutions=000020g}) is not unique: 
it belongs to the whole manifold of solutions those actions coincide.
As a result,  to obtain  physical quantities at low $\omega$ one should integrate over the whole saddle-point manifold, as it was done in 
Ref.~\cite{Skv-Krv} for Gaussian RP model.  General solution that belongs to the saddle-point manifold can be written in the form
\begin{equation}
g_{0}\left(\psi^{\dagger}\hat{T}^{\dagger}\hat{T}\psi,\psi^{\dagger}\hat{L}\psi\right)\equiv g_{T}\left(\psi^{\dagger}\psi,\psi^{\dagger}\hat{L}\psi\right)
\label{saddle-manifold}
\end{equation}
where $\hat{T}$ is the 4-dimensional supermatrix that rotates supervectors $\psi$ and $\psi^+$.
It obeys the symmetry property $\hat{T}^{\dagger}\hat{L}\hat{T}\equiv\hat{L}$.

In the next Section we will show that unique high-$\omega$  solution (\ref{eq:=000020saddle=000020solutions=000020g}) is applicable 
at $\omega \gg E_{Th} \sim \sqrt{\Delta \Gamma_0}$
while integration over saddle-point manifold (\ref{saddle-manifold}) can be employed at $\omega \ll \Gamma_0$.  Since we always have
$\Delta \ll \Gamma_0$, the combination of both approaches cover the whole range of frequencies we are interested in.

\section{Level correlation function: results and asymptotics}
\label{sec: answers}

In this section we calculate the DoS correlation function
and discuss its properties. The main expression for the correlation function
follows from \eqref{correlator=000020with=000020sources} and \eqref{eq:=000020Z=000020=00003D=000020int=000020dg=000020e(S=00005Bg=00005D)}:
\begin{equation}
R(E,\omega)=\frac{1}{2}+\frac{\Delta^{2}}{8\pi^{2}}\text{Re}\int D\left[g\right]\left[\frac{\partial^{2}S\left[g\right]}{\partial J_{A}\partial J_{R}}+\frac{\partial S\left[g\right]}{\partial J_{R}}\frac{\partial S\left[g\right]}{\partial J_{A}}\right]e^{S\left[g\right]}\biggr|_{J_{R},J_{A}=0}.\label{eq:=000020correlator=000020derivative=000020terms}
\end{equation}
where the action $S\left[g\right]$ is defined in Eq.(\ref{Action=000020generalized}).
 The expression above is still too complicated to evaluate it exactly for an arbitrary $\omega$, so we proceed by analyzing two complementary limits. First we consider  high-frequency regime, where functional integral is dominated by saddle-point solution~(\ref{s.p.=000020omega}); then we switch to the low-frequency regime, where integration over the full saddle-point manifold~(\ref{saddle-manifold}) is required. 

In the saddle-point approximation $g\left(\psi^{\dagger},\psi\right)$
should be substituted by the solution (\eqref{eq:=000020saddle=000020solutions=000020g}).
Quadratic over $g_\omega(\psi,\psi^+)$ term in the action does not depend on the sources $J_{A,R}$, it is also invariant under $\psi\rightarrow\hat{T}\psi$ transformations. Supersymmetry of this
term means that it does not contribute to the action on the
saddle-point manifold
(see integration theorems in Refs.~\cite{Mirlin:2000cla},\cite{Verbaarschot_2004} or  Supplementary material [\ref{C}]). The only important term left in the action is
\begin{equation}
S\left[g_{\text{s.p.}}\right]=N\ln\left\langle \int\left[d\psi\right]\exp\left(i\psi^{\dagger}\left(E\hat{L}+\frac{\Omega}{2}-\hat{J}\hat{K}\hat{L}-\zeta\hat{L}\right)\psi-g_{\text{s.p.}}\left(\psi,\psi^{\dagger}\right)\right)\right\rangle _{\zeta}
\end{equation}
where $g_{\text{s.p.}}$ is saddle-point solution of \eqref{s.p.=000020omega}.
Supersymmetric part of this expression should be equal to unity
and the other part is assumed to be small, so that one can expand the logarithm to get
\begin{equation}
S\left[g_{\text{s.p.}}\right]=N\left\langle \left\{ \int\left[d\psi\right]\exp\left(i\psi^{\dagger}\left(E\hat{L}+\frac{\Omega}{2}-\hat{J}\hat{K}\hat{L}-\zeta\hat{L}\right)\psi-g_{\text{s.p.}}\left(\psi,\psi^{\dagger}\right)\right)\right\} _{\hat{J},\hat{T}\neq0}\right\rangle _{\zeta}
\label{Sg4}
\end{equation}

Integration over $d\zeta$ in Eq.(\ref{Sg4} goes smoothly over
broad range of energies $\sim W$ which leads effectively to the restriction of $\psi^{\dagger}\hat{L}\psi $  being very small
(by the same logics as described in the analysis of the
saddle-point solution above). As a result, one can employ
$g_{\text{s.p.}}\left(\psi,\psi^{\dagger}\right)\biggr|_{\psi^{\dagger}\hat{L}\psi=0}$
from the solution \eqref{eq:=000020saddle=000020solutions=000020g} to get
\begin{equation}
S\left[g_{\text{s.p.}}\right]=N\left\langle \left\{ \int\left[d\psi\right]\exp\left(i\psi^{\dagger}\left(E\hat{L}+\frac{\Omega}{2}-\hat{J}\hat{K}\hat{L}-\zeta\hat{L}\right)\psi-g_{\text{s.p.}}\left(\psi,\psi^{\dagger}\right)\biggr|_{\psi^{\dagger}\hat{L}\psi=0}\right)\right\} _{\hat{J},\hat{T}\neq0}\right\rangle _{\zeta}\label{eq:=000020common=000020action}
\end{equation}

Further analysis differs for small and large $\omega$. First we consider the high-frequency region within saddle-point
approximation; the domain of applicability of these results becomes clear by comparison with results of exact calculation provided later for the low frequency region.

\subsection{High frequencies $\omega\gg E_{th}\equiv \sqrt{\Delta\Gamma/2\pi}$}

In the considered limit, correlations reflect the properties of the whole miniband, so fine structure is washed out and only the averaged properties matter. In this limit, the unique saddle-point solution $g_\omega(\psi^\dagger \psi,0)$ is sufficient.
In the high-$\omega$ limit  (parameter $\Gamma$ is defined in Sec.4.2) one employs $g_{\omega}\left(\psi^{\dagger}\psi,0\right)$ solution.
It is sufficient  to calculate  saddle-point action  as function of the sources:
\begin{equation}
S\left[g_{\omega}\right]=N\left\langle \left\{ \int\left[d\psi\right]\exp\left(i\psi^{\dagger}\left(E\hat{L}+\frac{\Omega}{2}-\hat{J}\hat{K}\hat{L}-\zeta\hat{L}\right)\psi-\left[\Gamma_{\omega}\psi^{\dagger}\psi\right]^{\mu/2}\right)\right\} _{\hat{J}\neq0}\right\rangle _{\zeta}
\end{equation}
Recalling properties of one-sided Lévy distribution and definition of  superdeterminant, we find
\begin{equation}
S\left[g_{\omega}\right]=N\int_{0}^{\infty}drL_{\frac{\mu}{2}}\left(r\right)\left\langle \left\{ \text{Sdet}^{-1}\left(E-\zeta+\left(\frac{\Omega}{2}+i\Gamma_{\omega}r\right)\hat{L}-\hat{J}\hat{K}\right)\right\} _{\hat{J}\neq0}\right\rangle _{\zeta}
\label{Sg2}
\end{equation}
At this stage it is useful to define the corresponding Green function
\begin{equation}
\hat{G}\equiv\left(E-\zeta+\left(\frac{\Omega}{2}+i\Gamma_{\omega}r\right)\hat{L}\right)^{-1}=\frac{E-\zeta-\left(\frac{\Omega}{2}+i\Gamma_{\omega}r\right)\hat{L}}{\left(E-\zeta\right)^{2}-\left(\frac{\Omega}{2}+i\Gamma_{\omega}r\right)^{2}}\label{eq:=000020hat=00007BG=00007D}
\end{equation}
Employing exact relation for superdeterminants, 
$\ln\text{Sdet}\hat{A}=\text{Str}\ln\hat{A}$
one can expand action in Eq.(\ref{Sg2}) it over sources $J_{R,A}$:
\begin{equation}\label{G def}
S\left[g_{\omega}\right]=N\int drL_{\frac{\mu}{2}}\left(r\right)\left[\left\langle \text{Str}\left[\hat{G}\hat{J}\hat{K}\right]\right\rangle _{\zeta}+\frac{1}{2}\left\langle \text{Str}^{2}\left[\hat{G}\hat{J}\hat{K}\right]\right\rangle _{\zeta}+\frac{1}{2}\left\langle \text{Str}\left[\hat{G}\hat{J}\hat{K}\hat{G}\hat{J}\hat{K}\right]\right\rangle _{\zeta}\right]
\end{equation}
Distribution $P_{D}\left(\zeta\right)$ is a very slow function of 
$\zeta$, as compared to $\zeta$-dependence of the Green function
$\hat{G}$  defined in \eqref{eq:=000020hat=00007BG=00007D},
so it is possible to use approximation $P_{D}\left(\zeta\right)\approx P_{D}\left(E\right)\sim W^{-1}$.
Performing integration near the pole  (with the use of the fact that
$\Im \Gamma_\omega < 0$) and also the relation $P_{D}\left(E\right)N=\Delta^{-1}$, we arrive at
\begin{equation}
S\left[g_{\omega}\right]=-i\frac{\pi}{\Delta}\left\{ 2\left(J_{R}-J_{A}\right)-8J_{R}J_{A}\int dr\frac{L_{\frac{\mu}{2}}\left(r\right)}{\Omega+2i\Gamma_{\omega}r}\right\} ,
\label{Sg3}
\end{equation}
Substitution of (\ref{Sg3}) into \eqref{eq:=000020correlator=000020derivative=000020terms}
gives final result in the form
\begin{equation}
\label{correlatorlargew}
R\left(\omega\right)=1+\frac{\Delta}{\pi}\int_{0}^{\infty}dr\frac{L_{\frac{\mu}{2}}\left(r\right)\cdot2r\text{Re}\Gamma_{\omega}}{\left[\omega-2r\text{Im}\Gamma_{\omega}\right]^{2}+\left[2r\text{Re}\Gamma_{\omega}\right]^{2}}
\end{equation}
In the limit $\mu \to 2$ the above result coincides with the one obtained in Ref.~\cite{Skv-Krv} for Gaussian RP model at high 
$\omega$.  For general $\mu$ similar result was obtained in Ref.~\cite{lunkin2024localdensitystatescorrelations} where local DoS correlation function $C(\omega)$ was obtained by means of cavity equation; the relation between these results is as follows:
$R(\omega) - 1 =  2^{\mu/2} \Delta \cdot C(\omega)$. The difference in numerical coefficients is due to slightly different models:
while we consider  Hermitian matrix ensemble with complex off-diagonal elements, the function $C(\omega)$ is calculated in 
~\cite{lunkin2024localdensitystatescorrelations} for real matrix
ensemble. At high frequencies the main asymptotics is given  by the power-law
\begin{equation}
    R(\omega) = 1 + \frac{\Delta}{\pi \Gamma_0} \frac{ 2^{\mu/2}\Gamma\left(\frac{\mu}{2}\right)\Gamma\left(\frac{\mu}{2}+1\right)}{ \Gamma\left(2 - \frac{2}{\mu}\right)} \left(\frac{\Gamma_0}{\omega}\right)^\mu
    \label{large w asymptotic}
\end{equation}
We present  details of this calculation 
in Appendix \ref{sec: mellin}.

\begin{figure}[!h]
    \centering
    \includegraphics[width=0.5\linewidth]{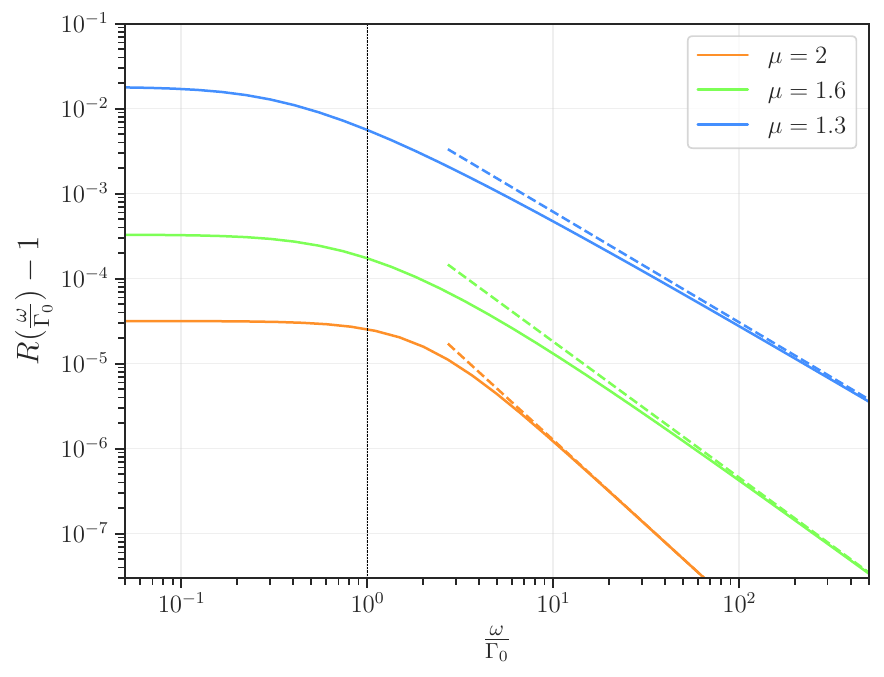}
    \caption{Correlation function $R(\omega) -1$ in a log-log scale, obtained from
    Eq.(\ref{correlatorlargew}) with $\Gamma_\omega$ found by means of numerical solution of Eq.(\ref{Gomega}). 
    We choose $\gamma = 1.1,\quad \sigma/W = 0.02,\quad N = 10^7$.  Dashed lines correspond to the asymptotic solution provided in Eq.\eqref{large w asymptotic}}.
    \label{fig:largefreq}
\end{figure}



\subsection{Domain $\omega\ll\Gamma_{0}$}

Now we can use expansion over parameter $\frac{\omega}{\Gamma_0} \ll 1$.
We will keep nonzero $\omega$ in the action 
\eqref{eq:=000020common=000020action} only
and replace $g_{\text{s.p.}}\left(\psi,\psi^{\dagger}\right)\biggr|_{\psi^{\dagger}\hat{L}\psi=0}$ used in the previous Sec.4.1 by
saddle-point manifold $g_{T}\left(\psi^{\dagger}\psi,0\right)$  
parametrized by the rotation matrix $\hat{T}$.
One should remember the definition $\omega+i0\equiv\Omega$, so that
if $\omega=0\text{, then }\Omega=i0$ to maintain the convergence of
integrals. 

At small energy differences,  the system resolves correlations within a single miniband. In this case, one must integrate over the full saddle-point manifold, which restores the characteristic Wigner–Dyson type oscillations at scales of the mean level spacing $\Delta$.

After inverse field transformation $\psi\rightarrow\hat{T}^{-1}\psi$
the action acquires the form 
\begin{equation}
S\left[g_{T}\right]=\frac{1}{\Delta}\int drL_{\frac{\mu}{2}}\left(r\right)\left\langle \left\{ \text{Sdet}^{-1}\left(E+\left(\frac{\Omega}{2}\hat{T}\hat{T}^{\dagger}+i\Gamma_{0}r\right)\hat{L}-\hat{T}\hat{J}\hat{K}\hat{L}\hat{T}^{\dagger}\hat{L}-\zeta\right)\right\} _{\hat{J},\hat{T}\neq0}\right\rangle _{\zeta}
\end{equation}
and integration over functions $D\left[g\right]$ in
\eqref{eq:=000020correlator=000020derivative=000020terms} is replaced
by the integration over $\hat{T}$ matrices. 
Matrix $\hat{T}$ is closely connected with Efetov matrix $\hat{Q}$
as $\hat{T}^{\dagger}\hat{T}=\hat{L}\hat{Q}$ (see Supplement \ref{sec: Efetov parametrization}).  
Further procedure is similar to the one used in the previous subsection. First of all one  performs an expansion over 
$\frac{\Omega}{\Gamma_{0}}$ and $\hat{J}$. Then, using the same tricks as in \eqref{G def}-\eqref{Sg3} one obtains an intermediate result in terms of $\hat{Q}$ matrices (remember that 
Str$\left[a \hat{1}+ b\hat{L}\right]$ = 0 for any numbers $a,b$).
\begin{equation}
S_{0}\left(\hat{T},\hat{J}\right)=\frac{i\pi}{\Delta}\text{Str}\left(\frac{\Omega}{2}\hat{L}\hat{Q}-\hat{J}\hat{K}\hat{Q}\right)+
\end{equation}
\[
\frac{\pi}{2\Delta\Gamma}\left\{ \text{Str}\left(\hat{J}^{2}-\Omega\hat{J}\hat{K}\right)-\text{Str}\left(\hat{J}\hat{K}\hat{Q}\hat{J}\hat{K}\hat{Q}-\Omega\hat{J}\hat{K}\hat{Q}\hat{L}\hat{Q}+\frac{\Omega^{2}}{4}\hat{L}\hat{Q}\hat{L}\hat{Q}\right)\right.+
\]
\[
\left.\text{Str}^{2}\left[\hat{J}\hat{K}\right]-\text{Str}^{2}\left(\hat{J}\hat{K}\hat{Q}-\frac{\Omega}{2}\hat{L}\hat{Q}\right)\right\}
\]

Finally, the key parameter $\Gamma$ is determined as follows:
\begin{equation}
\Gamma\equiv\left[\int dr\frac{L_{\mu/2}\left(r\right)}{2r\Gamma_{0}}\right]^{-1}=\frac{2\Gamma_{0}}{\Gamma\left(\frac{2}{\mu}+1\right)}.
\label{eq:=000020Gamma=000020skv=000020def}
\end{equation}
Using the relation $\hat{J}=J_{R}\frac{\hat{L}+1}{2}+J_{A}\frac{1-\hat{L}}{2}$, we calculate the derivatives
and obtain the following terms in the action \eqref{eq:=000020correlator=000020derivative=000020terms}  :
\begin{equation}
S\left[g_{T}\right]\biggr|_{J_{R,A}=0}=\frac{i\pi\Omega}{2\Delta}\text{Str}\left(\hat{L}\hat{Q}\right)-\frac{\pi\Omega^{2}}{8\Delta\Gamma}\left(\text{Str}\left(\hat{L}\hat{Q}\hat{L}\hat{Q}\right)+\text{Str}^{2}\left(\hat{L}\hat{Q}\right)\right)\label{eq:=000020S0(Q)}
\end{equation},

\begin{equation}
\frac{\partial^{2}S\left[g_{T}\right]}{\partial J_{A}\partial J_{R}}=\frac{\pi}{\Delta\Gamma}\left[\text{Str}\left(\hat{U}_{-}\hat{U}_{+}\right)+\text{Str}\left(\hat{U}_{-}\right)\text{Srt}\left(\hat{U}_{+}\right)+4\right]\label{eq:=000020d^2=000020S=000020(Q)}
\end{equation},

\begin{equation}
\frac{\partial S\left[g_{T}\right]}{\partial J_{R}}\frac{\partial S\left[g_{T}\right]}{\partial J_{A}}=-\left[\frac{i\pi}{\Delta}\text{Str}\left(\left[\hat{U}_{+}\right]\right)+\frac{\pi\Omega}{2\Delta\Gamma}\left(2-\text{Str}\left(\hat{U}_{+}\hat{L}\hat{Q}\right)-\text{Str}\left(\hat{U}_{+}\right)\text{Str}\left(\hat{L}\hat{Q}\right)\right)\right]\times\label{eq:=000020dS=000020dS=000020(Q)}
\end{equation}
\[
\left[\frac{i\pi}{\Delta}\text{Str}\left(\hat{U}_{-}\right)+\frac{\pi\Omega}{2\Delta\Gamma}\left(2-\text{Str}\left(\hat{U}_{-}\hat{L}\hat{Q}\right)-\text{Str}\left(\hat{U}_{-}\right)\text{Str}\left(\hat{L}\hat{Q}\right)\right)\right],
\]
where $\hat{U}_{+}\equiv\frac{\hat{L}+1}{2}\hat{K}\hat{Q}$ and $\hat{U}_{-}=\frac{\hat{L}-1}{2}\hat{K}\hat{Q}$.

The relation \eqref{eq:=000020Gamma=000020skv=000020def} above means that the quantity which should be averaged over Lévy distribution is the \textit{inverse} miniband width $1/r$, which is equivalent to the \textit{decay time} from the miniband.
Evaluation of integrals like the one present in 
Eq.(\ref{eq:=000020Gamma=000020skv=000020def}) is discussed in detail  in Ref. \cite{lunkin2024localdensitystatescorrelations}.
The quantity $\Gamma$ is similar to the one defined
in \cite{Skv-Krv} for the Gaussian RP model and coincides with it 
at $\mu=2$.

\begin{figure}[!h]
    \centering
    \includegraphics[width=0.9\linewidth]{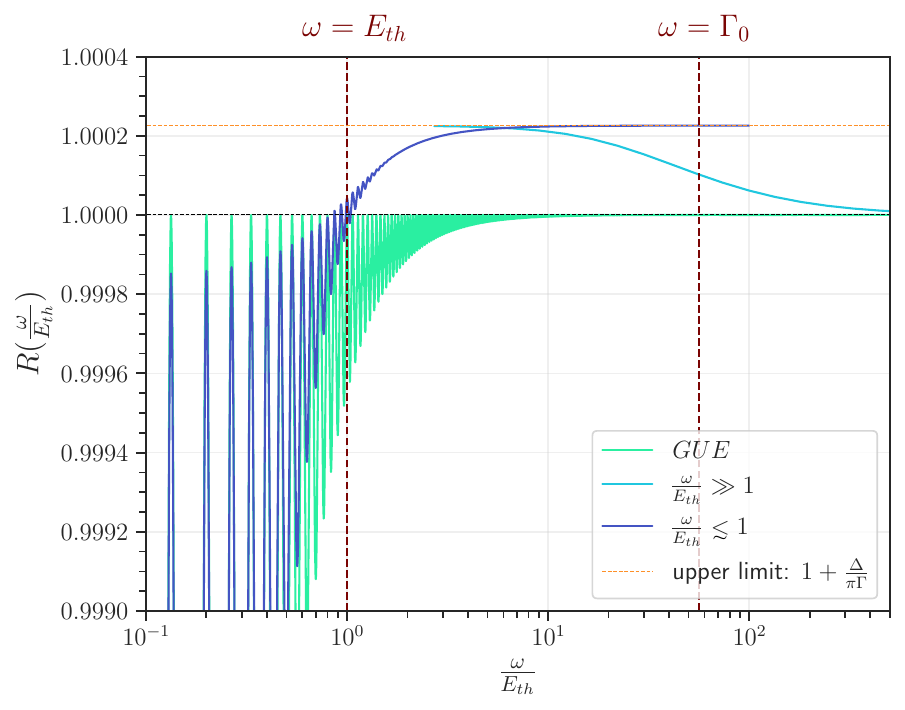}
    \caption{Level correlation function $R(\omega)$ obtained by means of  approximations \eqref{correlatorlargew} and \eqref{correlatormediumw}. Both approximations lead to nearly coinciding results at $\omega \approx E_{th}$.
    Here $\mu = 1.5,\quad \sqrt{\Delta / \Gamma_0} \approx 0.035, \quad\sigma \approx 0.023 W, \quad \gamma = 1.1,\quad N = 10^5$ (see Eq.\eqref{eq:=000020Gamma0=000020def}). Solutions \eqref{correlatorlargew} and \eqref{correlatormediumw} have an upper limit equal to $1 + \frac{\Delta}{\pi\Gamma}$. Contrary to the case of correlation function in the GUE ensemble, which never exceeds 1 
    (levels only repel each other), the
    L\'evy-RP model demonstrate weak long-range level attraction at 
    $\omega > E_{th}$.}
    \label{fig:correlationfunc}
\end{figure}

Now we  should integrate all manifold of $\hat{Q}$ in \eqref{eq:=000020d^2=000020S=000020(Q)}-\eqref{eq:=000020S0(Q)}.
 Unitary matrix $\hat{Q}$ is parameterized
in a standard way using Efetov parametrization (see Supplement \ref{sec: Efetov parametrization}). Two different energy scales appear in \eqref{eq:=000020S0(Q)}.
First term contains mean level spacing $\Delta$ and leads to the oscillations at $\omega\sim\Delta$, while the second one
defines Thouless energy $E_{th}\equiv\sqrt{\frac{\Delta\Gamma}{2\pi}}$, as an energy where typical GUE oscillations become exponentially suppressed. Combining all terms, we find the final integral expression for the correlation function at $\omega \ll \Gamma$:
\begin{multline}
R\left(E,\omega\right)=1+\frac{\Delta}{\pi\Gamma}+\\
\text{Re}\int_{1}^{\infty}d\lambda_{B}\int_{-1}^{1}d\lambda_{F}\left[\left(1+\frac{2i\Omega}{\Gamma}\lambda_{B}\right)^{2}+\frac{\Delta}{\pi\Gamma}\frac{\lambda_{B}}{\lambda_{B}-\lambda_{F}}\right]\exp\left(\frac{i\pi\Omega}{\Delta}\left(\lambda_{B}-\lambda_{F}\right)-\frac{\pi\Omega^{2}}{\Delta\Gamma}\lambda_{B}\left(\lambda_{B}-\lambda_{F}\right)\right)
\label{Rint}
\end{multline}
Double integral in Eq.(\ref{Rint}) can be further simplified using large parameter $\omega/\Delta \gg 1$ and we find
(see Supplement, Sec.\ref{sec:integrals} for details):
\begin{equation}
\label{correlatormediumw}
R\left(E,\omega\right)\approx1+\frac{\Delta}{\pi\Gamma}-\frac{\Delta^{2}}{2\pi^{2}\omega^{2}}\left(1-\cos\left(\frac{2\pi\omega}{\Delta}\right)\exp\left(-\frac{2\pi\omega^{2}}{\Delta\Gamma}\right)\right).
\end{equation}
The above result coincides with the one obtained for the Gaussian-RP model~\cite{Skv-Krv} up to renormalization of the miniband width $\Gamma$.
At low frequencies $\omega\ll E_{th}$ we get from Eq.(\ref{correlatormediumw}) a simplified expression
\begin{equation}
R\left(E,\omega\right)=1-\frac{\sin^{2}\left(\frac{\pi\omega}{\Delta}\right)}{\left(\frac{\pi\omega}{\Delta}\right)^{2}}+\frac{2\Delta}{\pi\Gamma}\sin^{2}\left(\frac{\pi\omega}{\Delta}\right),
\end{equation}
which coincides with GUE limit when $\Gamma/\Delta \to \infty$. 
The whole behavior of $R(\omega)$  at all frequencies is shown in Fig.\eqref{fig:correlationfunc}.

\section{Discussion and Conclusions}

We calculated energy level correlation function $R(\omega)$ for L\'evy Rosenzweig-Porter ensemble by means of supersymmetry method.
Our major new result is provided by Eqs.(\ref{correlatormediumw},
\ref{eq:=000020Gamma=000020skv=000020def}) refers to low-frequency range $\omega \leq E_{Th}$.  Functional form of 
Eq.(\ref{correlatormediumw}) reproduces the one known for Gaussian RP model~\cite{Skv-Krv}, while inverse of effective miniband width 
$1/\Gamma$ is given by the average over L\'evy distribution of local decay times, as follows from 
Eq.(\ref{eq:=000020Gamma=000020skv=000020def}).
In the high-frequency domain our result is given by Eqs.(\ref{correlatorlargew},\ref{Gomega}) and is in agreement with the
result of Ref.~\cite{lunkin2024localdensitystatescorrelations} for the correlation function of local density of states $C(\omega)$.

Both Gaussian RP and L\'evy-RP matrix ensembles share the same feature: at sufficiently small energy difference $\omega \ll E_{Th}$,
the level correlation function acquires the same form as in the usual GUE ensemble.  In the case of Gaussian RP model it is known
since Ref.~\cite{Skv-Krv} and it is interpreted in terms of miniband structure of energy levels. Indeed, the peculiar feature  of non-ergodic
phase in this type of model is that it becomes evident when  relatively large  energy window is considered, while narrow stripes of energy levels behave like in usual Wigner-Dyson matrix model. Our results demonstrate, surprisingly, that the same feature is retained even when
one allows for fat-tail distribution of matrix elements. However, for L\'evy-RP matrix ensemble the magnitude of the miniband width
$\Gamma$ and  of Thouless energy $ E_{Th} = \sqrt{\Delta \Gamma/2\pi}$
 should be calculated in the way different from the Gaussian RP case, see Eq.(\ref{eq:=000020Gamma=000020skv=000020def}).

The major qualitative difference between Gaussian RP and L\'evy-RP
ensembles is that the first one can be described in terms of
average Green functions $G(E)$ and self-energies $\Sigma(E)$, while in the L\'evy-RP case one is forced to consider non-trivial probability distributions for both Green function and self-energy.
Moreover, the miniband width $\Gamma_0$ known for Gaussian RP ensemble becomes a random quantity in the L\'evy-RP model, as can 
be observed with Eq.(\ref{eq:=000020Gamma=000020skv=000020def}):
effective width of a miniband $\Gamma$ is found to be an inverse
of a realization-dependent decay time $1/r\Gamma_0$ over L\'evy
distribution.  

Long power-law tail in the distribution of off-diagonal matrix elements makes minibands of L\'evy-RP ensemble different from
their Gaussian-RP counterparts which are compact in the values
of bare energies (diagonal matrix elements $\zeta_i$). Since there is a quite considerable probability to find abnormally large matrix element $H_{nm}$ in the L\'evy-RP case, here minibands are partially overlapping in the energy space.
Similar phenomena may be expected in other heavy-tail versions of the RP model, like the one studied in~\cite{Krav-Khaym}.

On a technical side, our results demonstrate that field-theoretic approach based on supersymmetry can be efficiently  employed for the analysis of systems described  by random Hamiltonian with heavy-tailed distributions.  We expect that such an approach might be useful for the analysis
of spatially extended systems with internal structure, similar to the one studied in Ref.~\cite{FyodorovMirlinSommers} but with a L\'evy distribution of hopping matrix elements.

We note  that our results justify  previous analyses performed in Refs.~\cite{BirTar_Levy-RP,Krav-Khaym,Monthus}. In these studies  the relation $I(N) \sim \frac{\Gamma_0}{N W}$ was used; it relates  typical scale of the inverse participation ratio  with the typical scale of imaginary part of  self-energy. We demonstrate by direct calculation that low-$\omega$ dynamics of the model indeed is equivalent to that of GUE, which puts the above assumption on 
firm ground. Our result Eq.(\ref{correlator=000020def}) for 
$R(\omega)$ correlation function implies GUE-type behavior of the spectral form factor $S(t)$  related to $R(\omega)$ by
Fourrier transform: $S(t)$ saturates at the value $I(N)$ for $t > t_H$,  where $t_H$ is the Heisenberg time.

\section{Acknowledgments} 

The authors are grateful to Yan Fyodorov, Vladimir Kravtsov, Pavel Ostrovsky and Marco Tarzia for many useful discussions.

\appendix

\section{Green functions and supersymmetric field theory}

\subsection{Supersymmetric field theory}

\label{A2} Representation of Green functions in supersymmetric field
theory is based on the property of determinants that
\begin{equation}
\ln\det\hat{A}=\text{Tr}\ln\hat{A}\Rightarrow\text{Tr}\left[\hat{A}\right]^{-1}=\frac{1}{2}\frac{\partial}{\partial J}\frac{\det\left(\hat{A}+J\right)}{\det\left(\hat{A}-J\right)}\biggr|_{J=0}.
\end{equation}

Whereas the $\left\langle \text{Tr}G_{R}\text{Tr}G_{A}\right\rangle $-type
function is represented as follows

\begin{multline}
\left\langle \text{Tr}\left[E+\frac{\Omega}{2}-\hat{H}\right]^{-1}\text{Tr}\left[E-\frac{\Omega}{2}-\hat{H}\right]^{-1}\right\rangle _{\hat{H}}=\\
=\frac{\partial^{2}}{\partial J_{R}\partial J_{A}}\left\langle \frac{\det\left(E-\hat{H}+\frac{\Omega}{2}+J_{R}\right)}{\det\left(E-\hat{H}+\frac{\Omega}{2}-J_{R}\right)}\frac{\det\left(E-\hat{H}-\frac{\Omega}{2}+J_{A}\right)}{\det\left(E-\hat{H}+\frac{\Omega}{2}-J_{A}\right)}\right\rangle _{\hat{H}}
\end{multline}

Using the basic properties of Gaussian integrals(for commutative and
anticommutative variables):

\[
\int e^{-\vec{\chi}^{\dagger}\hat{A}\vec{\chi}}d\vec{\chi}^{\dagger}d\vec{\chi}=\det\left(\frac{\hat{A}}{2\pi}\right),\quad\int e^{-\vec{S}^{\dagger}\hat{A}\vec{S}}d\vec{S}^{\dagger}d\vec{S}=\frac{1}{\det\left(\frac{\hat{A}}{2\pi}\right)}
\]

one arrives at the result \eqref{correlator=000020with=000020sources}. 

\textbf{Remark: }The sign of anticommutative variables does not matter
for the convergence of the integral; however, it is necessary choose the correct sign for the commuting variables.

\section{Derivation of \eqref{eq:=000020Z=000020=00003D=000020int=000020dg=000020e(S=00005Bg=00005D)},\eqref{Action=000020generalized}}

\label{B} 

\subsection{Averaging over off-diagonal matrix elements}

We start by averaging of the partition function (\ref{partitio=000020function})
over the random entries of $\hat{H}$ 
\begin{equation}
Z\left(E,\omega,\hat{J}\right)=\exp\left(i\sum_{n}^{N}\psi_{n}^{\dagger}\left[E\hat{L}+\frac{\Omega}{2}-\hat{J}\hat{K}\hat{L}\right]\psi_{n}\right)\times
\end{equation}
\[
\exp\left(\ln\left\langle \exp\left(-i\sum_{n,m}^{N}\psi_{n}^{\dagger}\left(\left[H_{L}\right]_{nm}+\delta_{nm}\left[H_{D}\right]_{nn}\right)\hat{L}\psi_{m}\right)\right\rangle _{\hat{H}_{L},\hat{H}_{D}}\right)
\]
The typical value of Gaussian elements is $\sim W$(assumed much larger than
Lévy diagonals typical value) so that it is reasonable to leave only
$\left[H_{D}\right]_{nm}$ on diagonal. This splits the averaging
$\left\langle ...\right\rangle $ into two independent parts. The
Hermitian property of the matrix $\hat{H}_{L}$ allows one to separate the
rest of sum into independent symmetrical entries, resulting in

\begin{equation}
Z\left(E,\omega,\hat{J}\right)=\left\langle \exp\left(i\sum_{n}^{N}\psi_{n}^{\dagger}\hat{L}\left[E+\frac{\Omega}{2}\hat{L}-\hat{J}\hat{K}-\left[H_{D}\right]_{nn}\right]\psi_{n}\right)\right\rangle _{\hat{H}_{D}}\times
\end{equation}
\[
\exp\left(\ln\left\langle \exp\left(-i\sum_{m<n}^{N}\left[\psi_{n}^{\dagger}\left[H_{L}\right]_{nm}\hat{L}\psi_{m}+\psi_{m}^{\dagger}\left[H_{L}\right]_{nm}^{*}\hat{L}\psi_{n}\right]\right)\right\rangle _{\hat{H}_{L}}\right)
\]
Since the symmetrical pairs of the matrix elements are independent,
the second line in the above equation can be rewritten as follows: 
\begin{equation}
\ln\left\langle \exp\left(-i\sum_{m<n}^{N}\left[\psi_{n}^{\dagger}\left[H_{L}\right]_{nm}\hat{L}\psi_{m}+\psi_{m}^{\dagger}\left[H_{L}\right]_{nm}^{*}\hat{L}\psi_{n}\right]\right)\right\rangle _{\hat{H}_{L}}=
\end{equation}
\[
=\frac{1}{2}\sum_{n\neq m}^{N}\ln\left\langle \exp\left(-i\left[\psi_{n}^{\dagger}\left[H_{L}\right]_{nm}\hat{L}\psi_{m}+\psi_{m}^{\dagger}\left[H_{L}\right]_{nm}^{*}\hat{L}\psi_{n}\right]\right)\right\rangle _{\hat{H}_{L}}.
\]
Furthermore, because there are $\sim N$ diagonal entries and $\sim N^{2}$ off-diagonal ones, one can replace
$\sum\limits_{m\neq m}$ by $\sum\limits_{m,n}$.
Later on, using the fact that off-diagonal matrix elements 
$\left[H_{L}\right]_{nm}$ are smaller that diagonal ones 
by the factor $N^{\gamma}$, one can use the following approximation
\begin{equation}
\sum_{n,m}^{N}\ln\left\langle \exp\left(-i\left[\psi_{n}^{\dagger}\left[H_{L}\right]_{nm}\hat{L}\psi_{m}+\psi_{m}^{\dagger}\left[H_{L}\right]_{nm}^{*}\hat{L}\psi_{n}\right]\right)\right\rangle _{\hat{H}_{L}}=\label{before=000020Levy=000020average}
\end{equation}
\[
\sum_{n,m}^{N}\ln\left[1+\left\langle \exp\left(-i\left[\psi_{n}^{\dagger}\left[H_{L}\right]_{nm}\hat{L}\psi_{m}+\psi_{m}^{\dagger}\left[H_{L}\right]_{nm}^{*}\hat{L}\psi_{n}\right]\right)-1\right\rangle _{\hat{H}_{L}}\right]\approx
\]
\[
\approx\sum_{n,m}^{N}\left\langle \exp\left(-i\left[\psi_{n}^{\dagger}\left[H_{L}\right]_{nm}\hat{L}\psi_{m}+\psi_{m}^{\dagger}\left[H_{L}\right]_{nm}^{*}\hat{L}\psi_{n}\right]\right)-1\right\rangle _{\hat{H}_{L}}\equiv\frac{1}{2N}\sum_{n,m}^{N}{\cal I}\left(\psi_{n}^{\dagger}\hat{L}\psi_{m}\right).
\]

Let us now denote $\left[H_{L}\right]_{nm}\equiv he^{i\theta}$ and $\psi_{n}^{\dagger}\hat{L}\psi_{m}\equiv t$, so that $\psi_{n}^{\dagger}\left[H_{L}\right]_{nm}\hat{L}\psi_{m}+\psi_{m}^{\dagger}\left[H_{L}\right]_{nm}^{*}\hat{L}\psi_{n}\equiv h\left(te^{i\theta}+t^{\dagger}e^{-i\theta}-i0\right)$,
where $-i0$ ensures convergence of the integral \ref{before=000020Levy=000020average}.
The object ${\cal I}(t)$ entering last line of Eq.(\ref{before=000020Levy=000020average}) can be rewritten as
\begin{equation}
{\cal I}(t)=2N\int\limits_{-\pi}^{\pi}\frac{d\theta}{2\pi}\int\limits_{0}^{\infty}\frac{d\left[h^{2}\right]}{2}P_{L}\left(h^{2}\right)\left(e^{-ih\left[te^{i\theta}+t^{\dagger}e^{-i\theta}-i0\right]}-1\right).
\end{equation}
Using normalization conditions, \ref{Levy=000020asymptotic} and following the calculations
in A.1 Appendix of \cite{lunkin2024localdensitystatescorrelations} paper one can proceed  to the following form:
\begin{equation}
{\cal I}(t)=\frac{2\sigma^{\mu}\Gamma(-\mu)}{N^{\gamma-1}\Gamma\left(-\frac{\mu}{2}\right)}\int\limits_{-\pi}^{\pi}\frac{d\theta}{2\pi}\left(i\left[e^{i\theta}t+e^{-i\theta}t^{\dagger}\right]+0\right)^{\mu},
\end{equation}
where constant follows from normalization.
To calculate the $\theta$ integral one can use its independence
on the phase of $t,t^{\dagger}$:
\begin{equation}
\int_{-\pi}^{\pi}\frac{d\theta}{2\pi}\left(i\left[e^{i\theta}t+e^{-i\theta}t^{\dagger}\right]+0\right)^{\mu}=\left|t\right|^{\mu}\int_{-\pi}^{\pi}\frac{d\theta}{2\pi}\left(0+2i\cos\theta\right)^{\mu}=\left|2t\right|^{\mu}\frac{\cos\left(\frac{\pi\mu}{2}\right)B\left(\frac{1}{2},\frac{1+\mu}{2}\right)}{\pi}
\end{equation}
Using the expression, one obtains the following result of the averaging over L\'evy distribution:
\begin{equation}
{\cal I}(t)=-\frac{\sigma^{\mu}|t|^{\mu}}{N^{\gamma-1}\Gamma\left(1+\frac{\mu}{2}\right)}
\end{equation}
As a result, we find
\begin{equation}
\ln\left\langle \exp\left(-i\sum_{n,m}^{N}\psi_{n}^{\dagger}\left[H_{L}\right]_{nm}\hat{L}\psi_{m}\right)\right\rangle _{\hat{H}_{L}}\approx-\frac{1}{2N}\sum_{n,m}^{N}\frac{\sigma^{\mu}\left[\psi_{n}^{\dagger}\hat{L}\psi_{m}\psi_{m}^{\dagger}\hat{L}\psi_{n}\right]^{\mu/2}}{N^{\gamma-1}\Gamma\left(1+\frac{\mu}{2}\right)}.
\end{equation}

\begin{equation}
Z\left(E,\omega,\hat{J}\right)=\label{Z=000020before=000020H-S}
\end{equation}
\[
\left\langle \int\left[d\psi\right]\exp\left(i\sum_{n}^{N}\psi_{n}^{\dagger}\hat{L}\left(E+\frac{\Omega}{2}\hat{L}-\hat{J}\hat{K}-\left[H_{D}\right]_{nn}\right)\psi_{n}-\frac{1}{2N}\sum_{n,m}^{N}\frac{\sigma^{\mu}\left[\psi_{n}^{\dagger}\hat{L}\psi_{m}\psi_{m}^{\dagger}\hat{L}\psi_{n}\right]^{\mu/2}}{N^{\gamma-1}\Gamma\left(1+\frac{\mu}{2}\right)}\right)\right\rangle _{\hat{H}_{D}}
\]

\subsection{Functional Hubbard-Stratonovich transformation}

An obvious difficulty that still remains is the non-analytic power
$\mu$ of $\psi_{n}^{\dagger}\hat{L}\psi_{m}\psi_{m}^{\dagger}\hat{L}\psi_{n}$
in the functional (instead of the quadratic term arising for the Gaussian distribution). This non-analyticity encodes the fat tails in the distribution which, in their turn, determine the peculiar physical properties of the system. In order to decouple the supervectors we use \textit{the functional Hubbard-Stratonovich(H-S) transformation}~\cite{MirFyod_non_Gauss} instead of the
usual one. Generalized expression looks as follows: 
\[
\exp\left(\frac{1}{2N}\int\left[d\psi\right]\left[d\psi'\right]v\left(\psi\right)C\left(\psi,\psi'\right)v\left(\psi'\right)\right)=
\]

\begin{equation}
\int Dg\exp\left(-\frac{N}{2}\int\left[d\psi\right]\left[d\psi'\right]g\left(\psi\right)C^{-1}\left(\psi,\psi'\right)g\left(\psi'\right)+\int\left[d\psi\right]g\left(\psi\right)v\left(\psi\right)\right),\label{<z>=000020before=000020saddle=000020point}
\end{equation}
where $C\left(\psi,\psi'\right)$, $v(\psi)$ and $g(\psi)$ are some
functions or fields.

The advantage of this method and its formal derivation was discussed
in detail in our previous paper~\cite{SafonovaLevyRPDoS} dedicated
to the calculation of the average DoS by the same method. 
Hence, only the final formulae will be provided in the present paper: 
\begin{equation}
\exp\left(-\frac{1}{2N}\cdot\frac{\sigma^{\mu}N^{1-\gamma}}{\Gamma\left(\frac{\mu}{2}+1\right)}\sum_{n,m}^{N}\left[\psi_{n}^{\dagger}\hat{L}\psi_{m}\psi_{m}^{\dagger}\hat{L}\psi_{n}\right]^{\mu/2}\right)=\label{result=000020of=000020H-S}
\end{equation}
\[
\int{\cal D}g\exp\left(\frac{N}{2}\int\left[d\psi\right]\left[d\psi'\right]g\left(\psi,\psi^{\dagger}\right)\left\{ \frac{\sigma^{\mu}N^{1-\gamma}}{\Gamma\left(\frac{\mu}{2}+1\right)}\left[\psi^{\dagger}\hat{L}\psi^{\prime}\psi^{\prime\dagger}\hat{L}\psi\right]^{\mu/2}\right\} ^{-1}g\left(\psi^{\prime},\psi^{\prime\dagger}\right)-Ng\left(\psi,\psi^{\dagger}\right)\right)
\]
Here we introduced functional integral over functions of superfields $g\left(\psi,\psi^{\dagger}\right)$.
Combining \ref{result=000020of=000020H-S} with
the previous expression (\ref{Z=000020before=000020H-S}) leads
to the equations (\ref{eq:=000020Z=000020=00003D=000020int=000020dg=000020e(S=00005Bg=00005D)},\ref{Action=000020generalized}) for the partition function.
Factor $N$ in the exponent comes due to $N$ independent integrations over $\psi_n,\psi_n^+$.

\section{Saddle-point equation and its solution}

\label{C}

\subsection{Derivation of the saddle-point equation}

Equating to zero variation of the action \eqref{Action=000020generalized} over $\delta g\left(\psi,\psi^{\dagger}\right)$, one obtains the following integral equation
for the saddle-point:
\begin{equation}
g_{\text{s.p.}}\left(\psi^{\prime\dagger},\psi^{\prime}\right)=\frac{\left\langle \int\left[d\psi\right]{\cal I}\left(\psi^{\prime\dagger}\hat{L}\psi\right)\exp\left(i\psi^{\dagger}\left(E\hat{L}-\zeta\hat{L}+\frac{\Omega}{2}\right)\psi-g_{\text{s.p.}}\left(\psi^{\dagger},\psi\right)\right)\right\rangle _{\zeta}}{\left\langle \int\left[d\psi\right]\exp\left(i\psi^{\dagger}\left(E\hat{L}-\zeta\hat{L}+\frac{\Omega}{2}\right)\psi-g_{\text{s.p.}}\left(\psi^{\dagger},\psi\right)\right)\right\rangle _{\zeta}}
\label{saddle-C}
\end{equation}

where ${\cal I}\left(x\right)\equiv\frac{\sigma^{\mu}N^{1-\gamma}}{\Gamma\left(\frac{\mu}{2}+1\right)}\left[x^{\dagger}x\right]^{\mu/2}$.  The structure of Eq.(\ref{saddle-C}) indicates that its solution is a function of two invariants:
$g_{\text{s.p.}}\left(\psi^{\dagger},\psi\right) = g_{\omega}\left(\psi^{\dagger}\psi,\psi^{\dagger}\hat{L}\psi\right)$.
Once we  search for the solution in this form,  the integrand of the
 integral in  the denominator is found to be invariant under the superunitary transformations
$\psi_{R,A}\rightarrow\hat{U}\psi_{R,A},\quad\psi=\left(\begin{array}{cc} \psi_{R} & \psi_{A}\end{array}\right)^{T}$, thus it is equal to unity. Therefore the final form of the saddle-point equation is
\begin{equation}
g_{\omega}\left(\psi^{\prime\dagger}\psi^{\prime},\psi^{\prime\dagger}\hat{L}\psi^{\prime}\right)=\left\langle \int\left[d\psi\right]{\cal I}\left(\psi^{\prime\dagger}\hat{L}\psi\right)\exp\left(i\psi^{\dagger}\left(E\hat{L}-\zeta\hat{L}+\frac{\Omega}{2}\right)\psi-g_{\omega}\left(\psi^{\dagger}\psi,\psi^{\dagger}\hat{L}\psi\right)\right)\right\rangle _{\zeta}.\label{eq:=000020g_=00005Comega=000020appendix=000020def}
\end{equation}
At $\Omega=0$ the saddle-point solution becomes
\begin{equation}
g_{\omega}\left(\psi^{\dagger}\psi,\psi^{\dagger}\hat{L}\psi\right)\biggr|_{\omega=0}\equiv g_{0}\left(\psi^{\dagger}\psi,\psi^{\dagger}\hat{L}\psi\right).
\end{equation}
Actually at $\Omega=0$ the whole saddle  manifold of solutions exist, which  can be parametrized by the rotation matrix $\hat{T}$
subject to the condition $ \hat{T}^{\dagger}\hat{L}\hat{T}=\hat{L}$:
\begin{equation}
\psi\rightarrow\hat{T}\psi,\quad g_{T}\left(\psi,\psi^{\dagger}\right)\equiv g_{0}\left(\psi^{\dagger}\hat{T}^{\dagger}\hat{T}\psi,\psi^{\dagger}\hat{L}\psi\right).
\end{equation}
Saddle-manifold solutions of this kind obey the equation
\begin{equation}
g_{T}\left(\psi^{\prime},\psi^{\prime\dagger}\right)=\left\langle \int\left[d\psi\right]{\cal I}\left(\psi^{\prime\dagger}\hat{L}\psi\right)\exp\left(i\psi^{\dagger}\left(E-\zeta\right)\hat{L}\psi-g_{T}\left(\psi,\psi^{\dagger}\right)\right)\right\rangle _{\zeta}.
\end{equation}

\subsection{Solution for the saddle-point equation}

Now our goal is to reduce 
Eq.\eqref{eq:=000020g_=00005Comega=000020appendix=000020def}
for a function of supervectors to simpler equations for functions of commuting variables. We use representation
\begin{equation}
\psi=\left(\begin{array}{cccc}
S_{R} & \chi_{R} & S_{A} & \chi_{A}^{*}\end{array}\right)^{T},\quad\psi^{\dagger}=\left(\begin{array}{cccc}
S_{R}^{*} & \chi_{R}^{*} & S_{A}^{*} & -\chi_{A}\end{array}\right)
\end{equation}
where $\frac{S_{R}}{S_{R}^{\prime}}=\frac{\left|S_{R}\right|}{\left|S_{R}^{\prime}\right|}e^{i\theta_{R}}$ 
and $\frac{S_{A}}{S_{A}^{\prime}}=\frac{\left|S_{A}\right|}{\left|S_{A}^{\prime}\right|}e^{i\theta_{A}}$, and we expand functions of supervectors over Grassmanian variables $\chi_R,\chi_A,\chi_R^*,\chi_A^*$.
It appears to be convenient to look for the solution as function of the arguments $\psi_{R}^{2}$ and $\psi_{A}^{2}$ and thus to introduce
a new function $\tilde{g}_{\omega}\left(\psi_{R}^{2},\psi_{A}^{2}\right) = g_{\omega}\left(\psi^{\dagger}\psi,\psi^{\dagger}\hat{L}\psi\right)$.
The expansion of an arbitrary function $f\left(\psi_{R}^{2},\psi_{A}^{2}\right)$ over its Grassmanian components
looks as follows:
\begin{center}
\begin{multline}
f\left(\psi_{R}^{2},\psi_{A}^{2}\right)=f\left(\left|S_{R}\right|^{2},\left|S_{A}\right|^{2}\right)+\chi_{R}^{*}\chi_{R}\frac{\partial f\left(\left|S_{R}\right|^{2},\left|S_{A}\right|^{2}\right)}{\partial\left[\left|S_{R}\right|^{2}\right]}+\\
\chi_{A}^{*}\chi_{A}\frac{\partial f\left(\left|S_{R}\right|^{2},\left|S_{A}\right|^{2}\right)}{\partial\left[\left|S_{A}\right|^{2}\right]}+\chi_{R}^{*}\chi_{R}\chi_{A}^{*}\chi_{A}\frac{\partial^{2}f\left(\left|S_{R}\right|^{2},\left|S_{A}\right|^{2}\right)}{\partial\left[\left|S_{R}\right|^{2}\right]\partial\left[\left|S_{A}\right|^{2}\right]}\label{eq:=000020grassmannians=000020expansion}
\end{multline}
\par\end{center}
To solve Eq.\eqref{eq:=000020g_=00005Comega=000020appendix=000020def}
one will need the last term  of the above equation only.
In these new coordinates, 
$\psi^{\dagger}\hat{L}\psi^{\prime}\psi^{\prime\dagger}\hat{L}\psi$
reads as follows:
\begin{equation}
\psi^{\dagger}\hat{L}\psi^{\prime}\psi^{\prime\dagger}\hat{L}\psi\stackrel{\chi_{R,A}=0}{=}\left|S_{R}\right|^{2}\left|S_{R}^{\prime}\right|^{2}+\left|S_{A}\right|^{2}\left|S_{A}^{\prime}\right|^{2}-2\left|S_{R}\right|\left|S_{R}^{\prime}\right|\left|S_{A}\right|\left|S_{A}^{\prime}\right|\cos\left(\theta_{R}-\theta_{A}\right)\geq0
\end{equation}
After integration over Grassmanian variables, 
Eq.\eqref{eq:=000020g_=00005Comega=000020appendix=000020def}
acquires  the form
\begin{multline}
\tilde{g}_{\omega}\left(\left|S_{R}^{\prime}\right|^{2},\left|S_{A}^{\prime}\right|^{2}\right)=\frac{\sigma^{\mu}N^{1-\gamma}}{\Gamma\left(\frac{\mu}{2}+1\right)}\times\int_{0}^{\infty}d\left|S_{A}\right|^{2}d\left|S_{R}\right|^{2}\\
\int_{0}^{2\pi}\frac{d\theta}{2\pi}\left[\left|S_{R}\right|^{2}\left|S_{R}^{\prime}\right|^{2}+\left|S_{A}\right|^{2}\left|S_{A}^{\prime}\right|^{2}-2\left|S_{R}\right|\left|S_{R}^{\prime}\right|\left|S_{A}\right|\left|S_{A}^{\prime}\right|\cos\theta\right]^{\frac{\mu}{2}}\times\\
\frac{\partial^{2}}{\partial\left[\left|S_{R}\right|^{2}\right]\partial\left[\left|S_{A}\right|^{2}\right]}\left\langle e^{i\left(E-\zeta+\frac{\Omega}{2}\right)\left|S_{R}\right|^{2}-i\left(E-\zeta-\frac{\Omega}{2}\right)\left|S_{A}\right|^{2}-g_{\omega}\left(\left|S_{R}\right|^{2},\left|S_{A}\right|^{2}\right)}\right\rangle _{\zeta}
\label{eq:=000020g_kappa=000020x-y}
\end{multline}
In principle, $\tilde{g}_{0}\left(\left|S_{R}^{\prime}\right|^{2},\left|S_{A}^{\prime}\right|^{2}\right)$ follows
from $\tilde{g}_{\omega}\left(\left|S_{R}^{\prime}\right|^{2},\left|S_{A}^{\prime}\right|^{2}\right)$.
In this case one should remember the definition 
$\omega+i0\equiv\Omega$, so that if $\omega=0\text{ than }\Omega=i0$ to maintain the convergence
in \eqref{eq:=000020g_=00005Comega=000020appendix=000020def}. 
For our purpose  the function $\tilde{g}_{\omega}\left(\left|S_{R}^{\prime}\right|^{2},\left|S_{R}^{\prime}\right|^{2}\right)$
is needed (it corresponds to $g_\omega(\psi^+\psi,0)$ in previous notations). \textcolor{black}{From this point one needs to proceed
with the analytical continuation assuming that $\mu>2$, to obtain
reasonable results.}  It can be calculated in a few steps:
\begin{enumerate}
\item Let us define the following function (in order to shorten few next  equations):
\begin{equation}
F\left(\left|S_{R}\right|^{2},\left|S_{A}\right|^{2}\right)=\int_{0}^{2\pi}\frac{d\theta}{2\pi}\left[\left|S_{R}\right|^{2}+\left|S_{A}\right|^{2}-2\left|S_{R}\right|\left|S_{A}\right|\cos\theta\right]^{\frac{\mu}{2}}
\end{equation}
with the property
\begin{equation}
\frac{\partial^{2}F\left(\left|S_{R}\right|^{2},\left|S_{A}\right|^{2}\right)}{\partial\left[\left|S_{R}\right|^{2}\right]\partial\left[\left|S_{A}\right|^{2}\right]}\biggr|_{\left|S_{R}\right|^{2}=\left|S_{A}\right|^{2}}=\frac{\left[\left|S_{R}\right|^{2}\right]^{\frac{\mu}{2}-2}}{\sqrt{\pi}}\frac{2^{\mu}}{4}\frac{\mu}{2}\frac{\Gamma\left(\frac{\mu-1}{2}\right)}{\Gamma\left(\frac{\mu}{2}-1\right)}
\end{equation}
and then integrate Eq.(\ref{eq:=000020g_kappa=000020x-y}) by parts:
\begin{multline}
\tilde{g}_{\omega}\left(\left|S_{R}^{\prime}\right|^{2},\left|S_{R}^{\prime}\right|^{2}\right)=\frac{\sigma^{\mu}N^{1-\gamma}}{\Gamma\left(\frac{\mu}{2}+1\right)}\left[\left|S_{R}^{\prime}\right|^{2}\right]^{\frac{\mu}{2}}\times\\
\left\{ \int_{0}^{\infty}d\left|S_{A}\right|^{2}\frac{\partial F}{\partial\left[\left|S_{A}\right|^{2}\right]}\biggr|_{\left|S_{R}\right|^{2}=0}\left\langle e^{-i\left(E-\zeta-\frac{\Omega}{2}\right)\left|S_{A}\right|^{2}-\tilde{g}_{\omega}\left(0,\left|S_{A}\right|^{2}\right)}\right\rangle _{\zeta}+\right.\\
+\int_{0}^{\infty}d\left|S_{R}\right|^{2}\frac{\partial F}{\partial\left[\left|S_{R}\right|^{2}\right]}\biggr|_{\left|S_{A}\right|^{2}=0}\left\langle e^{i\left(E-\zeta+\frac{\Omega}{2}\right)\left|S_{R}\right|^{2}-\tilde{g}_{\omega}\left(\left|S_{R}\right|^{2},0\right)}\right\rangle _{\zeta}+\\
\left.+\int_{0}^{\infty}d\left|S_{A}\right|^{2}d\left|S_{R}\right|^{2}\frac{\partial^{2}F\left(\left|S_{R}\right|^{2},\left|S_{A}\right|^{2}\right)}{\partial\left[\left|S_{R}\right|^{2}\right]\partial\left[\left|S_{A}\right|^{2}\right]}\left\langle e^{i\left(E-\zeta+\frac{\Omega}{2}\right)\left|S_{R}\right|^{2}-i\left(E-\zeta-\frac{\Omega}{2}\right)\left|S_{A}\right|^{2}-\tilde{g}_{\omega}\left(\left|S_{R}\right|^{2},\left|S_{A}\right|^{2}\right)}\right\rangle _{\zeta}\right\} \label{eq:=000020g_kappa=000020after=000020integration=000020by=000020parts}
\end{multline}
\item In case of smooth distribution one can use the following trick
\begin{equation}
\left\langle e^{i\left(E-\zeta\right)\left(\left|S_{R}\right|^{2}-\left|S_{A}\right|^{2}\right)}\right\rangle _{\zeta}\approx2\pi P_{D}\left(E\right)\delta\left(\left|S_{R}\right|^{2}-\left|S_{A}\right|^{2}\right)
\label{delta-f}
\end{equation}
so that \eqref{eq:=000020g_kappa=000020after=000020integration=000020by=000020parts}
reduces to
\begin{multline}
\tilde{g}_{\omega}\left(\left|S_{R}^{\prime}\right|^{2},\left|S_{R}^{\prime}\right|^{2}\right)=\frac{2^{\mu-1}\sigma^{\mu}N^{1-\gamma}P_{D}\left(E\right)}{\Gamma\left(\frac{\mu}{2}\right)\Gamma\left(\frac{\mu}{2}-1\right)}\Gamma\left(\frac{\mu-1}{2}\right)\left[\left|S_{R}^{\prime}\right|^{2}\right]^{\frac{\mu}{2}}\sqrt{\pi}\times\\
\int_{0}^{\infty}d\left[\left|S_{R}\right|^{2}\right]\left[\left|S_{R}\right|^{2}\right]^{\frac{\mu}{2}-2}e^{i\Omega\left|S_{R}\right|^{2}-\tilde{g}_{\omega}\left(\left|S_{R}\right|^{2},\left|S_{R}\right|^{2}\right)}
\end{multline}

Approximation (\ref{delta-f}) is valid if
$\left|\psi^\dagger \hat{L} \psi \right|$(equivalent to $|S_R|^2 - |S_A|^2$) is much larger than $\frac{1}{W}$. 
Using (\ref{eq:=000020independency=000020g}) and results from Ref.~\cite{SafonovaLevyRPDoS}, we estimate typical scale of
$g(\psi^\dagger \psi,\psi^\dagger \hat{L} \psi)$  as 
\begin{equation}
    g(0, \psi^\dagger \hat{L}\psi) \sim \frac{\sigma^\mu}{N^{\gamma - 1} W^{\mu/2}} \left|\psi^\dagger \hat{L} \psi \right|^{\mu/2} \Rightarrow \left|\psi^\dagger \hat{L} \psi \right| \sim \frac{N^{2\frac{\gamma - 1}{\mu}}W}{\sigma^2},
\end{equation}
which is indeed much larger than $1/W$.

\item Using the fact that $\tilde{g}_{\omega}\left(\left|S_{R}^{\prime}\right|^{2},\left|S_{R}^{\prime}\right|^{2}\right)=\left[\left|S_{R}^{\prime}\right|^{2}\Gamma_{\omega}\right]^{\mu/2}$,
one reduces the integral equation to the form of transcendental equation
\begin{multline}
\Gamma_{\omega}=\left[\frac{2^{\mu-1}\sigma^{\mu}N^{1-\gamma}P_{D}\left(E\right)}{\Gamma\left(\frac{\mu}{2}\right)\Gamma\left(\frac{\mu}{2}-1\right)}\Gamma\left(\frac{\mu-1}{2}\right)\sqrt{\pi}\int_{0}^{\infty}dxx^{\frac{\mu}{2}-2}e^{i\Omega x-\left[x\Gamma_{\omega}\right]^{\mu/2}}\right]^{\frac{2}{\mu}}=\\
=\left[\frac{2^{\mu-1}\sigma^{\mu}N^{1-\gamma}P_{D}\left(E\right)}{\Gamma\left(\frac{\mu}{2}\right)}\Gamma\left(\frac{\mu-1}{2}\right)\sqrt{\pi}\int_{0}^{\infty}drL_{\frac{\mu}{2}}\left(r\right)\left[-i\Omega+r\Gamma_{\omega}\right]^{1-\frac{\mu}{2}}\right]^{\frac{2}{\mu}}\label{eq:=000020g(x,=0000200)=000020final}
\end{multline}
which solves the saddle point equation \eqref{eq:=000020g_=00005Comega=000020appendix=000020def}
for any $\omega$. In particular, in the limit $\omega\rightarrow0+$ 
one obtains  the result \eqref{eq:=000020Gamma0=000020def}.
\end{enumerate}

\section{High frequencies asymptotics}\label{sec: mellin}
In this section we derive \eqref{large w asymptotic}. We need to use Mellin transform defined as
\begin{equation}
    {\cal M}_f(\lambda) \equiv \int_0^\infty dx f(x) x^{\lambda - 1}
\end{equation}
with the property
\begin{equation}
    \int_0^\infty dx f(x) g(x) = \int_{c-i\infty}^{c + i\infty} \frac{d
    \lambda}{2\pi i} {\cal M}_f(\lambda) {\cal M}_g(1-\lambda).
    \end{equation}

$c$ is the constant determined in a way that both ${\cal M}_f(\lambda)$ and ${\cal M}_g(1-\lambda)$ exist. Applying this to the integral in \eqref{Sg3} one receives precise expression

\begin{equation}
    \int_0^\infty dr \frac{L_{\mu/2}(r)}{\Omega + 2 i \Gamma_\omega r} =\frac{1}{2i\Gamma_\omega} \int_{c- i\infty}^{c + i\infty} \frac{d\lambda}{2\pi i} \frac{2}{\mu}\Gamma\left( \frac{2}{\mu} (1 - \lambda)\right) \Gamma(\lambda) \left( -\frac{i\Omega}{2\Gamma_\omega}\right)^{-\lambda},\quad 0<c<1.
\end{equation}

One can approximate it, counting only the nearest poles contribution $\lambda = 1, 1+\frac{\mu}{2}$. That gives

\begin{equation}
    \int_0^\infty dr \frac{L_{\mu/2}(r)}{\Omega + 2 i \Gamma_\omega r} \approx \frac{1}{2i\Gamma_\omega}\left[ \frac{2\Gamma_\omega}{-i\Omega} - \frac{\mu}{2} \left( \frac{2\Gamma_\omega}{-i\Omega} \right)^{\frac{\mu}{2} + 1}\right]
\end{equation}

After substituting this into \eqref{Sg3}, \eqref{eq:=000020correlator=000020derivative=000020terms} one will obtain \eqref{large w asymptotic} result. The same trick can be used to obtain second order approximations of \eqref{Gomega} and \eqref{correlatorlargew}.

\section{Efetov parameterization}\label{sec: Efetov parametrization}

Efetov parametrization for  4-dimensional supermatrix $\hat{Q}$ 
is defined as follows:
\begin{equation}
\hat{Q}\equiv\hat{T}^{-1}\hat{L}\hat{T}\equiv\left(\begin{array}{cc}
\hat{U}_{1} & 0\\
0 & \hat{U}_{2}
\end{array}\right)\hat{\Lambda}\left(\begin{array}{cc}
\hat{U}_{1}^{-1} & 0\\
0 & \hat{U}_{2}^{-1}
\end{array}\right),\quad\hat{\Lambda}=\left(\begin{array}{cccc}
\lambda_{B} & 0 & i\mu_{B} & 0\\
0 & \lambda_{F} & 0 & \mu_{F}^{*}\\
i\mu_{B}^{*} & 0 & -\lambda_{B} & 0\\
0 & \mu_{F} & 0 & -\lambda_{F}
\end{array}\right)
\end{equation}
Here $\hat{U}_{1,2}$ are Grassmannian matrices defined as
\begin{equation}
\hat{U}_{1}=\exp\left(\begin{array}{cc}
0 & -\alpha^{*}\\
\alpha & 0
\end{array}\right)=\left(\begin{array}{cc}
1-\frac{\alpha^{*}\alpha}{2} & -\alpha^{*}\\
\alpha & 1+\frac{\alpha^{*}\alpha}{2}
\end{array}\right),\enspace\hat{U}_{2}=\exp i\left(\begin{array}{cc}
0 & -\beta^{*}\\
\beta & 0
\end{array}\right)=\left(\begin{array}{cc}
1+\frac{\beta^{*}\beta}{2} & -i\beta^{*}\\
i\beta & 1-\frac{\beta^{*}\beta}{2}
\end{array}\right)
\end{equation}

\begin{equation}
\hat{U}_{1}^{-1}\left(\begin{array}{cc}
1 & 0\\
0 & -1
\end{array}\right)\hat{U}_{1}=\left(\begin{array}{cc}
1-2\alpha^{*}\alpha & -2\alpha^{*}\\
-2\alpha & -1-2\alpha^{*}\alpha
\end{array}\right),\enspace\hat{U}_{2}^{-1}\left(\begin{array}{cc}
1 & 0\\
0 & -1
\end{array}\right)\hat{U}_{2}=\left(\begin{array}{cc}
1+2\beta^{*}\beta & -2i\beta^{*}\\
-2i\beta & -1+2\beta^{*}\beta
\end{array}\right)
\end{equation}

and $\hat{\Lambda}$ contains the following commuting variables

\begin{equation}
\lambda_{B}=\cosh\theta_{B},\enspace\lambda_{F}=\cos\theta_{F},\enspace\mu_{B}=e^{i\phi_{B}}\sinh\theta_{B},\enspace\mu_{F}=e^{i\phi_{F}}\sin\theta_{F},\enspace
\end{equation}

\[
\text{Constraints}=\begin{cases}
0\leq\theta_{B}<\infty,\quad1\leq\lambda_{B}<\infty\\
0\leq\theta_{B}\leq\pi,\quad-1\leq\lambda_{F}\leq1\\
0\leq\phi_{B,F}\leq2\pi
\end{cases}
\]
with the following relations
\begin{equation}
\left|\mu_{B}\right|^{2}=\lambda_{B}^{2}-1,\quad\left|\mu_{F}\right|^{2}=1-\lambda_{F}^{2}
\end{equation}

Mesure of integration over Efetov matrix $\hat{Q}$ reads as
\begin{equation}
d\hat{Q}=-\frac{d\lambda_{B}d\lambda_{F}d\phi_{B}d\phi_{F}}{\left(\lambda_{B}-\lambda_{F}\right)^{2}}d\alpha d\alpha^{*}d\beta d\beta^{*}
\end{equation}

\section{Evaluation of the integral in 
Eq.\eqref{Rint}}\label{sec:integrals}

Is this section we provide details on how we  obtained the result shown in Eq.\eqref{correlatormediumw}. The starting point is the integral in \eqref{Rint}. Since the large parameter is $\kappa = \frac{\pi \omega}{\Delta} \gg 1$, one needs to obtain an answer up to the first order in $1/\kappa \ll 1$. If $\omega \sim E_{th}$ then $\kappa \frac{\omega}{\Gamma} \sim 1$ so that it is reasonable to denote $\frac{\omega}{\Gamma}$ as $\frac{p}{\kappa},\quad p\sim 1$. With  these notations, integral in \eqref{Rint} will take the form

\begin{equation}\label{Y = Y1 + Y2}
    {\cal Y} = {\cal Y}_1 + \frac{p}{\kappa^2} {\cal Y}_2
\end{equation}

\begin{equation}
{\cal Y}_1=\frac{1}{2}\int_{1}^{\infty}d\lambda_{B}\int_{-1}^{1}d\lambda_{F}\left(1+2i\frac{p}{\kappa}\lambda_{B}\right)^{2}e^{i\kappa\left(\lambda_{B}-\lambda_{F}\right)-p\lambda_{B}\left(\lambda_{B}-\lambda_{F}\right)} =
\end{equation}
\[
\frac{1}{2i\kappa}\int_{1}^{\infty}d\lambda_{B}\frac{\left(1+2i\frac{p}{\kappa}\lambda_{B}\right)^{2}}{1 + i \frac{p}{\kappa} \lambda_B}e^{i\kappa\left(\lambda_{B}-1\right)\left(1 + i \frac{p}{\kappa} \lambda_B\right)} \left(e^{2i\kappa\left(1 + i \frac{p}{\kappa} \lambda_B\right)} -1\right)
\]

\begin{equation}
    {\cal Y}_2=\int_{1}^{\infty}d\lambda_{B}\int_{-1}^{1}d\lambda_{F}\frac{\lambda_{B}}{\lambda_{B}-\lambda_{F}}e^{i\kappa\left(\lambda_{B}-\lambda_{F}\right)-p\lambda_{B}\left(\lambda_{B}-\lambda_{F}\right)}
\end{equation}

${\cal Y}_1$ is easily integrated over $\lambda_F$, while ${\cal Y}_2$ requires an additional step. Let us use
\begin{equation}
    \frac{d {\cal Y}_2}{d p} = -\int_{1}^{\infty}d\lambda_{B}\int_{-1}^{1}d\lambda_{F}\lambda_{B}^2 e^{i\kappa\left(\lambda_{B}-\lambda_{F}\right)-p\lambda_{B}\left(\lambda_{B}-\lambda_{F}\right)} =
\end{equation}
\[
\frac{i}{\kappa}\int_{1}^{\infty}d\lambda_{B}\frac{\lambda_B^{2}}{1 + i \frac{p}{\kappa} \lambda_B}e^{i\kappa\left(\lambda_{B}-1\right)\left(1 + i \frac{p}{\kappa} \lambda_B\right)} \left(e^{2i\kappa\left(1 + i \frac{p}{\kappa} \lambda_B\right)} -1\right)
\]

Both integrals collects on the $\lambda_B - 1 < \frac{1}{\kappa}$ scale so that one can make $\lambda_B = 1 + \frac{x}{\kappa}$ substitution and then expand over $\frac{1}{\kappa}$ parameter up to the lowest order. Note that $\Omega \equiv \omega +i0$ maintains the convergence.

\begin{equation}
    {\cal Y}_1 \approx \int_0^\infty dx \frac{i e^{(i-0)x}}{2 \kappa^2} \left(1 - e^{2i\kappa - 2p}\right) = \frac{1}{2\kappa^2} \left(  e^{2i\kappa - 2p} - 1 \right)
\end{equation}

\begin{equation}
     \frac{d {\cal Y}_2}{d p} = \frac{i}{\kappa^2} \int_0^\infty dx  e^{(i-0)x}\left( e^{2i\kappa - 2p} - 1 \right) = \frac{1}{\kappa^2}\left(1 - e^{2i\kappa - 2p}\right) \Rightarrow  {\cal Y}_2 = \frac{2p + e^{2 i \kappa - 2 p}}{2\kappa^2} + \text{const}
\end{equation}

To restore the constant we apply $p=0$. This integral is easily evaluated after its derivation over $\kappa$ and later integration. Constant can be found in $\kappa \rightarrow \infty$ limit.

\begin{equation}
   {\cal Y}_2\biggr|_{p=0} = \frac{i}{\kappa} + \frac{e^{2 i \kappa} - 1}{2\kappa^2} \Rightarrow {\cal Y}_2 = \frac{e^{2 i \kappa - 2 p} - 1 + 2p}{2\kappa^2} + \frac{i}{\kappa} .
\end{equation}

As one can see from \eqref{Y = Y1 + Y2}, the second term lowest order is much smaller so it is enough to consider ${\cal Y} \approx {\cal Y}_1$ only. Having restored all the notations, one should obtain the final expression \eqref{correlatormediumw}.

\bibliography{Levy-RP}
\end{document}